\documentclass{article}

\usepackage[english]{babel}
\usepackage{comment}
\usepackage[letterpaper,top=2cm,bottom=2cm,left=3cm,right=3cm,marginparwidth=1.75cm]{geometry}
\usepackage[affil-it]{authblk} 
\usepackage{etoolbox}
\usepackage{lmodern}

\makeatletter
\patchcmd{\@maketitle}{\LARGE \@title}{\fontsize{16}{19.2}\selectfont\@title}{}{}
\makeatother

\usepackage{graphicx}
\usepackage[colorlinks=true, allcolors=blue]{hyperref}
\usepackage{amsmath,bm}
\usepackage{siunitx}
\usepackage{wrapfig}
\usepackage{authblk}

\newcommand{\pbar}{\mbox{$\overline{\mathrm p}$}}
\newcommand{\Hbar}{\mbox{$\overline{\mathrm H}$}}
\newcommand{\nuHF}{$\nu_{\mathrm{HF}}$}

\newcommand{\nupone}{$\nu_{\pi_1}$}
\newcommand{\nuptwo}{$\nu_{\pi_2}$}

\title{Experimental perspectives on the matter-antimatter asymmetry puzzle: developments in electron EDM and $\bar{\textrm{H}}$ experiments}

\author[1]{D.~Comparat}
\author[2,3]{C.~Malbrunot}
\author[2,4]{S.~Malbrunot-Ettenauer}
\author[5]{E.~Widmann}
\author[6]{P.~Yzombard}
\affil[1]{Laboratoire Aim\'e Cotton, CNRS, Universit\'e Paris-Sud, ENS Paris Saclay, Universit\'e Paris-Saclay, Bâtiment 505, 91405 Orsay, France}
\affil[2]{TRIUMF, Vancouver, British Columbia V6T 2A3, Canada}
\affil[3]{McGill University, Montr\'eal, Québec, H3A 2T8, Canada}
\affil[4]{Department of Physics, University of Toronto, Toronto M5S 1A7, Canada}
\affil[5]{Stefan Meyer Institute for Subatomic Physics, Austrian Academy of Sciences, 1030 Vienna, Austria}
\affil[6]{Laboratoire Kastler Brossel, Sorbonne Universit\'e, CNRS, ENS-PSL Universit\'e, Coll\`ege de France, Paris, 75252, France}

\begin{document}
\maketitle

\begin{abstract}
In the search for clues to the matter-antimatter puzzle, experiments with atoms or molecules play a particular role. These systems allow measurements with very high precision, as demonstrated by the unprecedented limits down to $10^{-30}$ e.cm on electron EDM using molecular ions, and relative measurements at the level of $10^{-12}$ in spectroscopy of antihydrogen atoms. Building on these impressive measurements, new experimental directions offer potentials for drastic improvements. We review here some of the new perspectives in those fields and their associated prospects for new physics searches.
\end{abstract}

\section{Introduction}

The  asymmetry between the matter and antimatter content of our universe is one of the most tantalizing puzzle of modern physics. Many experimental endeavours are focused on searching for manifestations of the mechanisms that would produce such an asymmetry. The main contender for an asymmetry-generating mechanism is CP (Charge and Parity) symmetry violation which, together with the other Sakharov conditions \cite{Sakharov:1967dj}, would generate  baryon asymmetry. In the Standard Model CP-violation present in the CKM matrix fall short by about 10 orders of magnitude to be a possible explanation for the observed baryon asymmetry of the universe (BAU). The observation of BAU is thus one of the clear indications of the existence of physics beyond the Standard Model (SM). The possibility  of a baryon asymmetry generation through heavy neutron leptons (HNL) creation in the early universe has triggered a renewed interest in HNL searches at colliders (see for example \cite{Agrawal_2021}, references therein and numerous dedicated work). Direct measurements of CP violation in the lepton sector are carried out by long-baseline neutrino experiments \cite{Agarwalla2023, doi:10.1146/annurev-nucl-102014-021939}.
In a complementary approach to searches at high energies, low energy precision experiments are looking for 
lepton-number violating processes such as neutrinoless double beta decay \cite{doi:10.1146/annurev-nucl-101918-023407} or scrutinize atoms and molecules in the quest for the existence of electric dipole moments (EDM) of elementary particles that would indicating P and T (time-reversal) symmetry violations \cite{safronova2018search} the latter being equivalent to CP violation within a CPT-conserving theory. Many beyond-the-Standard-Model theories predict EDMs close to current experimental sensitivities \cite{alarcon2022electric}. 

In this paper we will review some of the  new experimental perspectives for EDM searches using atoms and molecules which have potential for impressive improvements to the state-of-the-art while facing challenges that are currently being addressed by a new and active community.  
Alternatively, BAU can be addressed through the lens of CPT violation. CPT violation combined with baryon number  violation could lead to baryon asymmetry  in thermal equilibrium and would have the advantage of being otherwise independent of C- and CP-violating processes \cite{RevModPhys.53.1, COHEN1988913}. However, since CPT conservation, through the so-called CPT theorem, is established as a cornerstone of quantum field theory, theories involving spontaneous breaking of CPT symmetry are much less common. Nevertheless, CPT violation arises in various string theory models (e.g. \cite{Alan_Kosteleck__1996, Ellis_2013}), as well as in models where the assumptions of the CPT theorem, such as Lorentz invariance or unitarity are not fulfilled. The group of Kostelecky has constructed an effective field theory, the Standard Model Extension (SME), a generalization of the Standard Model and General Relativity that allows for arbitrary Lorentz and CPT violations. Those violations are quantitatively controlled by a set of coefficients whose values are to be determined or constrained by experiment \cite{PhysRevD.55.6760}. Tests of CPT violation through the comparison of fundamental properties of matter and antimatter particles have been pursued in many sectors; from collider experiments to quasi table-top experiments. In particular, since about 20 years, experiments at the antiproton decelerator (AD) at CERN have been dedicated to taming antihydrogen atoms in order to allow extremely precise comparisons of their atomic spectral transitions with the analog ones measured in hydrogen. A precision of $2\times 10^{-12}$ \cite{Ahmadi:2018a} has been reached on the 1S-2S transition of trapped antihydrogen atoms which is only 3 orders of magnitude away from the best precision achieved on the same transition in a beam of hydrogen. Triggered in part by the successful laser cooling of antihydrogen atoms \cite{Baker:2021}, perspectives for substantial improvement of  precision are being investigated with slow beams of hydrogen in a fountain setting.
Aside from spectroscopy measurements, experiments at the AD started more recently exploring diverse experimental techniques to measure the effect of gravity on neutral atoms made - as far as valence quarks are concerned - of antimatter. As with spectroscopy measurements, the realization of sources of slow (anti)-hydrogen atoms will be the path to high precision. In particular an exciting  perspective lies in the first measurement of (anti)-atomic gravitational quantum states.

\section{Directions in  next generation EDM measurements}
An EDM is originating from a spatial separation of opposite electrical charge. If an EDM of elementary particle exists this extra moment should be aligned with the spin of the particle to avoid creating an extra degree of freedom that would make two different electrons, for example, discernible in violation of the Pauli principle.  Since the spin changes sign under time reversal while the spatial charge distribution remain unchanged, an EDM of an elementary particle would be an indication of T (and thus through CPT theorem of CP) violation.  Such EDMs of elementary particles clearly differ from conventional EDMs such as the ones present in polar molecules which are not violating time reversal symmetry.
As discussed in many recent review articles \cite{blum2022fundamental,alarcon2022electric,boeschoten2023perspectives},
the quest for an EDM of a fundamental particle originated over seventy years ago with Ramsey and Purcell's proposal to test discrete symmetries like parity. 
EDMs of elementary particles are strongly suppressed in the Standard Model of particle physics but are predicted with different magnitudes in various new physics scenarios. 
%This idea laid the foundation for  tests of various theoretical frameworks, notably the Standard Model of particle physics.
As time passed, numerous experimentally viable systems, including atoms and molecules, have contributed to setting increasingly stringent constraints on potential EDMs.
%In this context, we explore the perspectives of these experiments concerning statistics, systematics, and sensitivities to physics beyond the Standard Model.

Studying both paramagnetic and diamagnetic systems is necessary for probing different CP violating mechanisms that induce EDMs.
Here, we'll first focus on paramagnetic systems, particularly electron EDM (eEDM), which use methods similar to those for diamagnetic systems \cite{safronova2018search, alarcon2022electric}.
In gas phase experiments, spectroscopy techniques measure the energy shift $\Delta E = -\bm d_e.\bm E_{\rm eff} - \bm \mu.\bm B$ due to applied external electric $\bm E_{\rm ext}$ and magnetic $\bm B$ fields. 
The relationship between $\bm E_{\rm ext}$ and the effective field $\bm E_{\rm eff}$ seen by the electron is connected to Schiff's theorem and relativistic effects  yielding to \cite{Fukuyama2012} 
$E_{\rm eff} \sim (\qty{200}{GV/cm})(Z/80)^3{{\epsilon}}_s{{\epsilon}}_p$ 
for a $|s{\rangle }$ (we use the standard notation for the orbital angular  momentum) valence electron wave-function $\left|{\Psi }\right\rangle {\approx }{{\epsilon }}_s\left|s\right\rangle +{{\epsilon }}_p\left|p\right\rangle +{\dots }$ (value typically reduced by an order of magnitude for a $d$ electron or two orders of magnitude for an $f$ one) with a $\left|p\right\rangle $ 
character created by the external Stark Shift.  
Thus, high-$Z$ atoms %, most of them being radioactive,  
 with strongly s-p hybridized electron wavefunctions under an external field are ideal.
For atoms, weak polarization by $\bm E_{\rm ext}$ leads to $E_{\rm eff} \propto \bm E_{\rm ext}$ in the range of $\qtyrange{1e4}{1e7}{V/cm}$,
while polar molecules require only a small $E_{\rm ext}$ (few V/cm) for full polarization (Stark mixing of opposite parity states) of the molecular bond, giving $\epsilon_s \sim \epsilon_p \sim 1/\sqrt{2}$ and thus many orders of magnitude higher effective field, in the  $\qtyrange{1e10}{1e11}{V/cm}$ range.

\subsection{EDM searches with radioactive molecules}

%As just explained  radioactive molecules are ideal for eEDM measurement.

%In the longer term, combining the Schiff moment enhancement of deformed nuclei with relativistic enhancements in heavy polar molecules could lead to sensitivity orders of magnitude beyond current possibilities, opening new avenues for CP-violating physics searches.

%An essential area of technological development is the use of deformed nuclei for EDM searches, as they can lead to significantly larger atomic and molecular EDMs due to enhanced nuclear Schiff moments \cite{behr2022nuclei}.
%3.	Radioactive species (from \cite{roussy2022new})

Molecules in which one of the constituent atoms contains a radioactive nucleus represent intriguing systems for EDM searches. As explained above, high-Z atoms provide a substantial increase in sensitivity for eEDM experiments. The ACME collaboration, for instance, exploits the naturally occurring thorium isotope $^{232}$Th ($Z=90$ and a half-life of $1.4\cdot10^{10}$~years) incorporated in a thorium monoxide (ThO) molecule. Taking advantage of an internal electric field of $E_{\rm eff}=78$~GV/cm in the used excited electronic state, ACME currently establishes a limit on the eEDM of $|d_e| < 1.1\cdot10^{-29}$~e~cm (90\% confidence limit) \cite{ACME2018}. This compares favourably to $E_{\rm eff}=23$~GV/cm found in singly-charged hafnium monofluoride (HfF$^+$) which builds upon stable Hf atoms with $Z=72$. While its HfF$^+$ measurement holds 
since recently the tightest bound on an eEDM, i.e. $|d_e| <4.1\cdot10^{-30}$~e~cm \cite{HfF_eEDM2023,PhysRevA.108.012804}, the next generation JILA experiment will replace HfF$^+$ with ThF$^+$. This is motivated by the higher $Z$ in (radioactive) Th and, thus, larger $E_{\rm eff}$ as well as a longer coherence time reflecting that the EDM sensitivity is found in the electronic ground state of the molecule \cite{PhysRevA.105.022823}. 

Inspired by the successes and prospects for molecules with long-lived radionuclides, the research with molecules has recently been extended to molecular systems containing short-lived atomic nuclei \cite{RadMoleculesWhitePaper2023}. In these cases the half-lives are as short as weeks, days or even less. Hence, these radioisotopes do not occur naturally but can be synthesized in specialized radioactive ion beam (RIB) facilities. Note that the radioactivity itself is in most cases not of advantage for these EDM studies but rather an unavoidable feature of specific nuclides of interest, e.g. when extending to higher Z or selecting specific nuclear shapes all atomic nuclei are eventually found to be radioactive.  Thus, exploiting short-lived radionuclides allows one to tailor molecular systems ideally suited for next-generation EDM searches. For instance, atomic nuclei along an isotopic chain of a chemical element possess  different nuclear spins which provides an opportunity to systematically explore nuclear-spin dependency of EDM-inducing electron-nucleon couplings within the same molecular system. Such a program would be difficult to pursue when being limited to stable or long-lived nuclides. In the previously mentioned molecules, thorium has a single long-lived isotope, $^{232}$Th with a nuclear spin $I=0$ and is in addition to an eEDM sensitive to nuclear spin-independent electron-nucleon interaction \cite{PhysRevC.91.035502}.  The same is true for most stable or long-lived hafnium isotopes. The exceptions, $^{177}$Hf ($I=7/2$) and $^{199}$Hf ($I=9/2$), exhibit large spins which complicates experimental efforts given the many states induced by the hyperfine interaction. 

As another major advantage, the work with short-lived radioactive globally opens the region ‘above’ the stable lead isotope $^{208}$Pb on the nuclear chart in which no chemical element has a stable isotope and many also lack long-lived, naturally occurring ones. For example, the longest-lived francium (Fr) isotope is $^{223}$Fr with a half-life as short as 23 min. However, intriguing physics phenomena manifests in this region in and around the actinides. In particular, strong evidence for octupole-deformed ('pear-shaped') nuclei has been collected in radon (Rn), Fr, radium (Ra), actinium (Ac), Th, and protactinium (Pa) isotopes \cite{JBehr_octupoleDeformed2022,RadMoleculesWhitePaper2023} (and references therein). This exquisite shape, rarely observed in atomic nuclei, facilitates an enhanced sensitivity for time-reversal violation inside the atomic nucleus which would otherwise - by force of a 're-arrangement' of the electron cloud- be strongly obscured from experimenters probing atomic or molecular EDMs. This shielding effect \cite{PhysRev.132.2194} is formulated in Schiff's theorem in the limit of nonrelativistic quantum mechanics but is incomplete also due to the finite size and deformation of the atomic nucleus. This is referred to as the  nuclear Schiff moment. As the nucleus interacts with the surrounding electrons, the Schiff moment causes a polarisation of the electron cloud which leads to an atomic or molecular EDM. This gives rise to small energy shifts in atomic/molecular states when the spin of the atomic nucleus is aligned along, or against, an external electric field that polarizes the electron cloud. 

For octuple-deformed nuclei the Schiff moment can be large such that the sensitivity of detecting nuclear T-reversal breaking via an atomic or molecular EDM is significantly enhanced compared to spherical nuclei \cite{PhysRevLett.76.4316}. At this time, the tightest experimental bound on a nuclear Schiff moment arises from measurements in the (stable) mercury isotope $^{199}$Hg. Its measured EDM limit of $|d_{Hg}|<7.4\cdot10^{-30}$ e~cm (95\% confidence limit) translates to a nuclear Schiff moment of $|S_{Hg}|<3.1\cdot10^{-13}$ e~fm$^3$ \cite{PhysRevLett.116.161601}. Octuple-deformed nuclei are estimated to come along with enhancement factors for nuclear Schiff moments which can be two to three orders of magnitude larger than in $^{199}$Hg. Experimental efforts to directly benefit from this increased sensitivity are pursued in Ra, e.g. \cite{PhysRevLett.114.233002}. Another strategy, newly adopted for the next generation of experiments, is to complement the advantages offered by heavy, octupole-deformed nuclei with the additional enhancement factors when incorporating them into molecules \cite{RadMoleculesWhitePaper2023}. These can be as large as 1’000-10'0000 resulting to a combined increase in intrinsic sensitivity to nuclear Schiff moments by up to seven orders of magnitude compared to $^{199}$Hg. Such an enormous discovery potential has its price; all experimentally known examples of (static) octupole deformation are found in radioactive nuclei and are consequently available only in minute sample sizes at RIB facilities. 

Low yields and short half-lives are commonly encountered in the science of rare isotopes. Thus, a sophisticated tool box of experimental techniques has been developed to address them. Among those, methods of atomic, molecular and optical (AMO) physics such as ion and atom trapping as well as laser spectroscopy have been specifically adapted to the challenges when exploiting physics opportunities with RIBs \cite{Blaum_2013}. 
Exemplary highlights under 'extreme experimental conditions' include ion-trap based mass measurements of radionuclides with half-lives below 10~ms \cite{PhysRevLett.101.202501} or with yields less than 1 ion per hour delivered to experimenters \cite{doi:10.1126/science.1225636}. Laser spectroscopy of radioactive atoms has a long history of successful exploitation of high-resolution methods such as collinear laser spectroscopy as well as highly sensitive approaches such as resonant ionization spectroscopy. The field is continuously re-inventing itself by exploring new experimental methods as well as by expanding its scientific horizon \cite{YANG2023104005}.

As a ground-breaking example of this innovative processes, the successful formation and identification of molecules containing Ra isotopes in an ion trap \cite{PhysRevLett.126.023002} as well as the first spectroscopy of radium monofluoride (RaF) molecules \cite{GarciaRuiz2020,PhysRevLett.127.033001} have been the starting whistle for experimental work with radioactive molecules. While the general perspective of this newly emerging field is summarized in Ref.~\cite{RadMoleculesWhitePaper2023}, molecules which have been proposed in the context of EDM searches are for instance  RaF \cite{PhysRevA.82.052521,PhysRevA.90.052513,GarciaRuiz2020}, RaOH \cite{Isaev_2017,PhysRevLett.119.133002}, RaO \cite{PhysRevA.77.024501}, RaH \cite{PhysRevA.99.052502}, RaOCH$_3^+$ \cite{PhysRevLett.126.023002,PhysRevLett.126.023003}, AcF \cite{10.1063/5.0159888}, or short-lived ThO \cite{PhysRevC.99.035501}, as well as ThOH$^+$, ThF$^+$ \cite{PhysRevC.99.035501} or AcO$^+$ \cite{PhysRevA.101.042504}. 

Although their intrinsic enhancement factors are enormous, the radioactive nature of octupole-deformed nuclides will generally translate into lower statistics compared to stable molecules or atoms. Hence, experimental initiatives to measure nuclear Schiff moments with radioactive molecules often consider atom and ion traps as well as embedding them inside cryogenic solids to extend the coherence and observation time of the available samples. Moreover, studies have been initiated to optimize the efficient formation of these molecules \cite{AU2023375}, to obtain colder samples, e.g. via direct laser cooling, sympathetic cooling with co-trapped laser-coolable ions \cite{PhysRevResearch.4.033229} (or atoms), or formation of molecules starting from ultra-cold atoms themselves \cite{Klos_2022}. Since present knowledge about the structure of radioactive molecules is in most cases based on theoretical calculations, one critical experimental effort will focus on the spectroscopy of these molecules in order to further characterize their suitability for EDM searches. 

The emerging interest in (short-lived) radioactive molecules in general and in their potential for unprecedented sensitivity in EDM searches in particular coincides with the completion of next-generation RIB facilities such as FRIB at MSU, ARIEL at TRIUMF, FAIR at GSI, or SPIRAL2 at GANIL. An increased variety in RIBs is therefore expected for experiments, in combination with a significantly larger availability of specific radionuclides of interest. The latter will be facilitated by new multi-user capabilities either via parasitic use of radioactive ions downstream of another RIB experiment or by multiple driver beams at a single RIB facility enabling simultaneous RIB production and delivery to various experimental stations. Since known octupole deformed nuclei have with a few exceptions half-lives of days to years, offline operation of a previously irradiated production target or isotope harvesting from a facility’s beam dump are attractive resources of radionuclides of interest. Hence, the worldwide network of RIB facilities provides an excellent foundation for the research of radioactive molecules and their use as novel EDM probes \cite{RadMoleculesWhitePaper2023}. 

\subsection{Solid state systems and cryogenic matrices}

\begin{figure}[h]
\begin{center}
    \includegraphics[width=\textwidth]{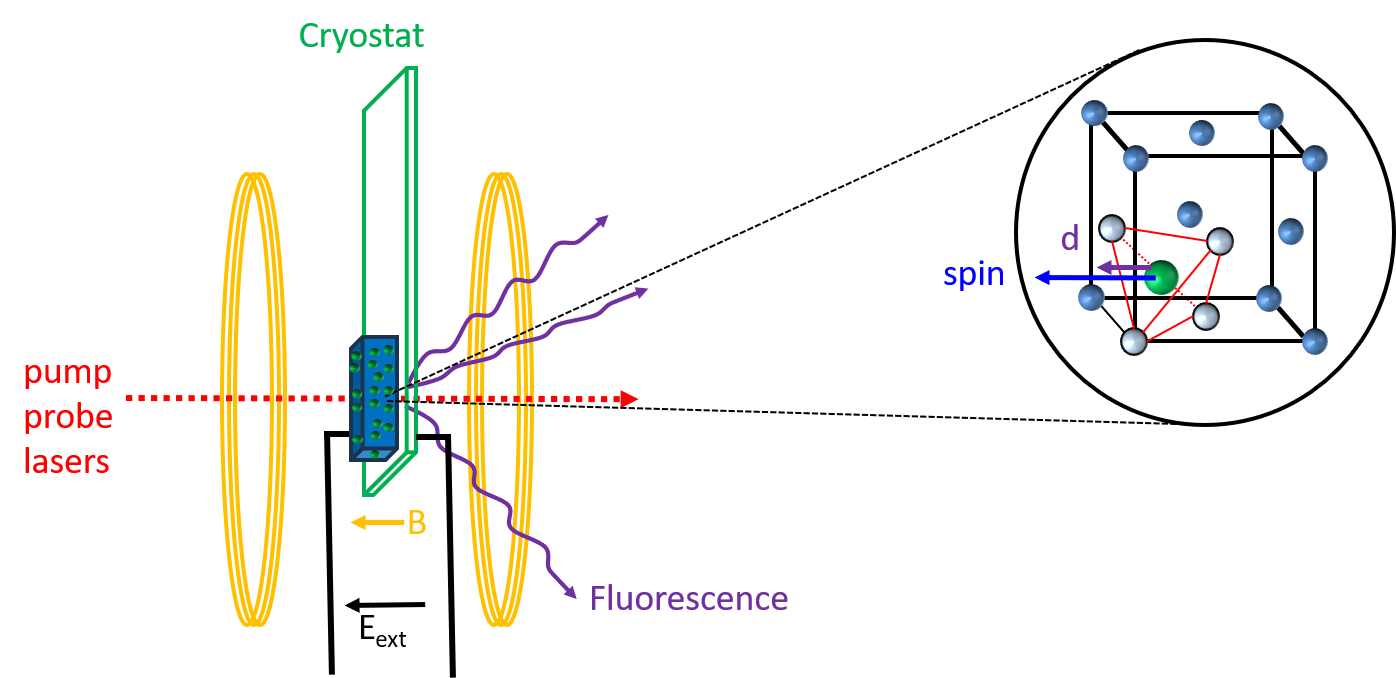} 
    \caption{Schematics of a solid state EDM experiment. A solid (in blue), often formed at or cooled to cryogenic temperature, is doped with high density atoms or molecules (in green), pumped and probed by lasers under external electric and/or magnetic fields. The 
    alignment of  the EDM $\bm d$ along the spin is schematized on the zoomed part on the right: illustrative case for Cs, or BaF, doped in solid Ar  \cite{battard2023cesium,lambo2023calculation}
    with 
 a trapping site with one dopant surrounded by 4 vacancies (white) in a tetrahedral  symmetry of an fcc crystal (in blue).}
\label{fig:EDM}
\end{center}
\end{figure}

The measurement of permanent EDM in atomic, molecular, and nuclear systems can be conducted in both the gas phase and solid phase. While the gas phase technique is the most used and established, solid samples can offer advantages. The first key edge being
the macroscopic number $N$ of atoms or molecules present in a given sample during the total integrated interrogation time directly improving the statistical averaging of the 
sensitivity $\sigma_d$ of an EDM measurement: 
$
\sigma_d \approx \frac{\hbar }{ \epsilon E_{\rm eff}\tau \sqrt{N}}
$; 
%is given by the time-energy relation with
 where ${\tau}$ is the single-shot (travelling or coherence time  limited) measurement time, and ${\epsilon}$  the visibility or polarization fraction of atoms/molecules (i.e. dipole alignment or spin state preparation efficiency). So, by combining the very high atomic density with the large sample size of solids,  many order of magnitude improvement in sensitivity can theoretically be expected compared to beam or ion trap experiments.
The second advantage, suggested by Shapiro in 1968, is that new observables, in addition to the  pure spectroscopy shift $\Delta E$, can be used. 
This is due to the fact that the electric dipole moment $\bm d$ of a particle is aligned to its total angular momentum and thus to its magnetic moment $\bm \mu$. Compared to the gas phase, 
two additional measurement schemes can be used:  alignment of $\bm d$ using $\bm E_{ext}$, creating macroscopic magnetization due to $\bm \mu \| \bm d$; or alignement of $\bm \mu$ using $\bm B$, leading to dielectric polarization (since $\bm d  \| \bm \mu$). Both effects can be measured in a sample.
A typical setup for such solid state experiment is shown in Figure \ref{fig:EDM}.

\subsubsection{EDM in solid: results to date}

\begin{table}[h!]\footnotesize\hspace{-10pt}
\begin{tabular}{|l|l|l|l|l|}
\hline
year & sample                         & eEDM limit in e·cm                                                  & method              &          \\ \hline
1961 & KCr- \& NH$_4$- (SO$_4$)$_2$ 12H$_2$O  & $d_e < 10^{-13}$                                                    & spectroscopy    & \cite{browne1961effects}  \\ \hline
1963 & Al$_{2}$O$_3$:Cr \& MgO:Cr        & $d_e = (1 \pm 4.6)× 10^{-16}$                                    &  spectroscopy    & \cite{royce1963linear} \\ \hline
1979 & nickel-zink-ferrite            & $d_e = (8.1 \pm 11.6)× 10^{-23}$                                 & Magnetometry        & {\cite{vasil1978measurement}} \\ \hline
2004 & GdIG                           & $d_e = (2 \pm 3)× 10^{-24}$                                      & Voltage  & \cite{heidenreich2005limit} \\ \hline
2011 & GGG                            & $d_e = (-5.57 \pm 7.98 \pm 0.12)× 10^{-25}$  & Magnetometry        & \cite{kim2015new} \\ \hline
2012 & Eu$_{0.5}$Ba$_{0.5}$TiO$_3$            & $d_e = (-1.07 \pm 3.06 \pm 1.74)× 10^{-25}$ & Magnetometry        & \cite{eckel2012limit} \\ \hline
\end{tabular}
\caption{List of solid state systems used for eEDM searches and their published bound results. Whenever stated, statistical and systematic uncertainties are provided in this order.}
\label{table:eEDM in solid history}
\end{table}

We have listed in Table \ref{table:eEDM in solid history} all (to our knowledge)  eEDM in solid state results. The first successful eEDM search in a solid system was performed  in 1961 on a mono-crystal at cryogenic temperatures using a modified version of the electron paramagnetic resonance (EPR) spectroscopy. The measurement employed an electric field to measure the linear stark splitting due to the eEDM. In 1963, an improved measurement using Cr atoms in alumina or magnesia applied both electric and magnetic fields, enabling a line shift measurement. However, after these works, no further eEDM search using EPR in solids was published.
 In 1979, a measurement was performed on nickel-zinc ferrite  using an oscillating electric field and a SQUID to measure the magnetization from the eEDM-induced sample polarization. Later,  two new magnetometry measurements were conducted on paramagnetic Gadolinium-Gallium-Garnet (GGG) and Eu$_{0.5}$Ba$_{0.5}$TiO$_3$. The GGG measurement had improved sensitivity and systematics control, while the Eu solid  experiment faced challenges due to ferroelectricity and ferroelectric relaxation. Despite the difficulties, the Eu solid  measurement provided the most stringent eEDM limit in a solid-state experiment to date due to the enhanced polarizability inside the ferroelectric crystal.
Only one attempt was made to measure the eEDM-induced electric potential after applying a magnetic field. The experiment used ferrimagnetic GdIG with Yttrium-doped samples to achieve specific compensation temperatures. However, complications arose from the pinning of magnetic domains, significantly limiting the measurement's sensitivity.

\subsubsection{EDM in solid: new directions}

Following previous work in the above-mentioned experiments, a new direction started to recently gain importance: the use of atoms or molecules embedded in a frozen gas at cryogenic temperatures.
The idea of matrix isolation spectroscopy to measure the eEDM is not new and was first proposed by C. Pryor and F. Wilczek in 1987 \cite{pryor1987artificial}. The use of molecules was then mentioned in Refs. \cite{hinds1991testing,kozlov2006proposal}.
 The main idea is to isolate EDM-relevant species in rare gas crystals, which are chemically inert elements and thus should not affect strongly their hosts, allowing them to be manipulated using lasers similarly to molecular beam experiments. 
One of the first serious experimental investigations was undertaken by the group of A. Weis (see the review \cite{moroshkin2008atomic}) using  laser desorption of Cs  atoms in pressurized solid $^4$He.
Results were not  reproducible because of ion and electrons created by the laser shots with the clear-cut conclusion that "\textit{one cannot expect an experiment using cesium atoms implanted in helium crystals to be competitive with other existing EDM experiments.}" \cite{ulzega2006theoretical}.  
However, thanks to the work of 
 J. D. Weinstein's group, on alkali atoms in para-hydrogen, Ne, and Ar matrix \cite{BarnesEtAl1981,BondybeyEtAl1996,Crepin-GilbertTramer1999} there has been renewed interest in using  matrix isolation spectroscopy for EDM searches \cite{HfF_eEDM2023}. Indeed, recent achievements regarding high dopant density $\sim \qty{1e18}{cm^3}$, while maintaining long spin coherence time (as long as \qty{0.1}{s}) and good optical pumping capacity, are all very positive first steps toward an initial EDM measurement or spin alignment \cite{kanagin2013optical,upadhyay2019spin,upadhyay2020ultralong,dargyte2021optical}. 
 After the proposal \cite{vutha2018oriented}, a series of  studies on atoms (Rb, Cs, Tm, Yb, ...) and molecules (BaF, RaF, ..) in rare-gas (Ne,Ar mainly) or parahydrogen matrices were conducted\cite{lambo2021high,braggio2022spectroscopy,gaire2022excitation,battard2023cesium,li2023optical,lambo2023calculation,lambo2023calculationNe,ballof2023progress,azevedo2023platform}.
Besides the high sample size and other advantages mentioned above, measurements in solid can also offer a few potential additional benefits:  the crystal field (typically on the order of tens of MV/cm) can produce high $E_{\rm eff}$ values,  molecules could align themselves along the local crystal axes due to specific trapping site environment \cite{vutha2018oriented,lambo2023calculation,lambo2023calculationNe}. This phenomenon would mitigate many of the systematic effects associated with applying an external electric field during the EDM measurement cycle. Obviously many other specific systematic effects have to be mastered along the way to an EDM measurement including laser and fields stability alignment and homogeneity, isotopic purity quality and reproductibility of the samples, inhomogeneous broadening effects, magneto-electric coupling, bleaching effect,  diffusion of particles, residual magnetization or polarization,  dielectric relaxation, heat formation or lattice deformation under application of fields or cryostat-induced vibrations etc.

% Brian Odom, NOrthwestern:   growing single-crystal samples 

% Giovanni Carugno, University of Padova / INFN. + Marco Guarise INFN Ferrara Caterina Braggio University of Padova R. Calabrese  \cite{braggio2022spectroscopy}

% Daniel Comparat, Laboratoire Aimé Cotton works on Cs in Ar \cite{battard2023cesium}

% Colin Parker, Georgia Institute of Technology: magnetic dipole transition of the Tm atoms trapped in solid argon and neon 

% Eric Hessels, York University. BaF in Ar or Ne \cite{lambo2023calculation,lambo2023calculationNe}

% Jaideep Singh, Michigan State University. The under-construction  Facility for Rare Isotope Beams  FRIB-EDM -instrument RaF in Ar \cite{ballof2023progress}. 
 
% time-reversal symmetry violation using Pa-229 ions trapped in optical crystals \cite{singh2019new}

%  Nuclear T-violation search using octupole-deformed nuclei in a crystal \cite{ramachandran2023nuclear}

% Amar Vutha BaF in Ne \cite{li2023optical}

% Y. Xie, A. A. Buchachenko, J. T. Singh High Resolution Spectroscopy of Neutral Yb Atoms in a Solid Ne Matrix \cite{lambo2021high} 

% Claudio Cesar, Instituto de Fisica - UFRJ's group: Study Matrix isolation sublimation for producing cryogenic beams of atoms and molecules \cite{sacramento2015matrix} that can also be a platform for trapped cryogenic electrons, anions and cations for fundamental physics \cite{azevedo2023platform}.

These will open and feed collateral areas of research such as  sensitive (co-)magnetometers for instance with co-dopants or using polyatomic molecules  \cite{anderegg2023quantum}, single crystals growth \cite{SwansonEtAl1986,guarise2020feasibility,bhandari2021high}, individual atomic detection by optical means or by ionisation and electrons’ extraction \cite{nexo2019imaging,guarise2020particle,guarise2022particle},   
quantum control and  sensing
\cite{budker2022quantum} that may help in reaching higher sensitivities in the future.

\subsubsection{Related experiments}

The EDM experimental apparatus  can often be used to detect other spin-dependent interactions or  other tiny effects that would affect the atomic properties \cite{kimball2023probing}. Among them is 
the search for ultralight bosonic dark matter \cite{jackson2023search} such as axion like particles that typically produced 
 time-varying EDMs, the effect of which
 being possibly detected using atomic transitions \cite{sikivie2014axion,braggioaxion}. This 
opens new perspectives for dark matter searches such as followed by the 
Cosmic Axion Spin Precession Experiment (CASPEr)  \cite{budker2014proposal}
or by the
DEMIURGOS project \cite{guarise2020particle}. The nEXO experiment is another example of the application of trapping in cryogenic matrices. nEXO is searching for neutrinoless double beta decay in xenon;  the tagging of the $^{136}$Ba atom daughter, resulting from the decay of $^{136}$Xe, is a powerful technique to discriminate background. One of the techniques for Ba-tagging consists of trapping the atom in a solid xenon matrix and detecting it through fluorescence imaging \cite{nexo2019imaging}. % The nEXO collaboration has achieved the imaging of individual barium atoms in solid xenon on a sapphire window
 Similar EDM type of solid state experiments can sometimes be realized using liquid, where spin precession (of NMR type) measurement can be favored compared to solid phase, but keeping the advantage of the high density 
 \cite{kuchler2014novel,wu2019search}.
However in solid, the existence of a preferential axis of symmetry allow for anisotrope couplings between angular momenta. In such systems new parity violating observable arise giving sensitivity to nuclear anapole moments  for instance
\cite{leggett1977macroscopic,leggett1978macroscopic,flambaum1992long,bouchiat2001atomic,mukhamedjanov2005manifestations}.

Ultimately, this solid state methodology can be effectively integrated with radioactive atoms or molecules housed within crystals. A number of experiments are progressing towards identifying the best conditions using a transparent crystal, ideal for optical pumping, which is either doped or stoichiometric for maximum density, and based on radioactive molecules, which provide high enhancement and co-magnetometry properties, along with a closed, separately aligned parity doublet \cite{singh2019new,ballof2023progress,ramachandran2023nuclear,morris2023rare,RadMoleculesWhitePaper2023,sushkov2023effective}.
This illustrates the profusion of new developments in this area and presages exciting results.

\section{Directions in next generation $\bar{\textrm{H}}$ experiments}

\subsection{Toward fountain experiments }
%\input{texs/Antihydrogen_spectroscopy}
%% EW
\label{spectrofountain}

\subsubsection{State of the art}

\begin{figure}[h]
\begin{center}
    \includegraphics[width=70mm]{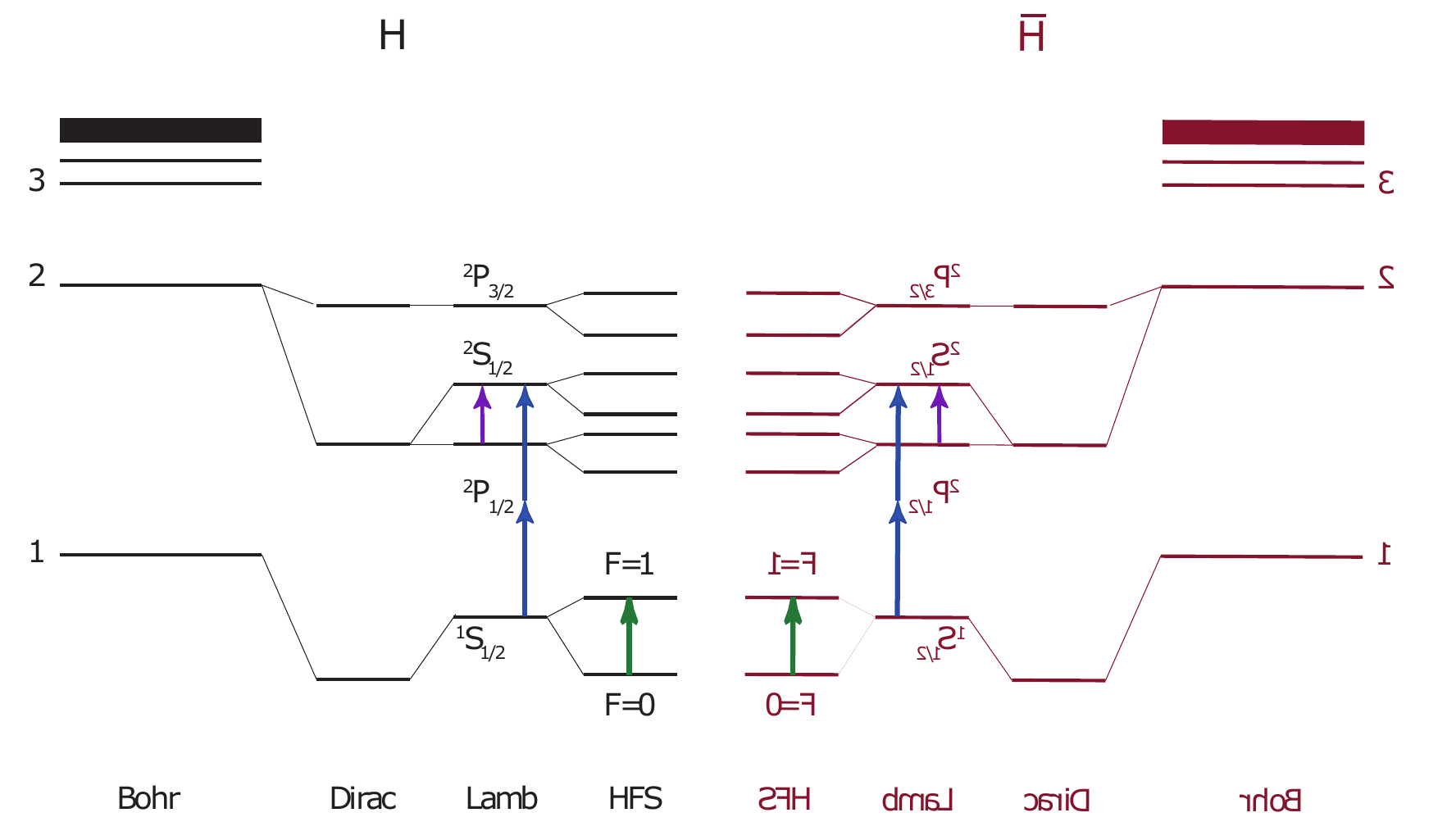} \hspace*{5mm}
    \includegraphics[width=45mm]{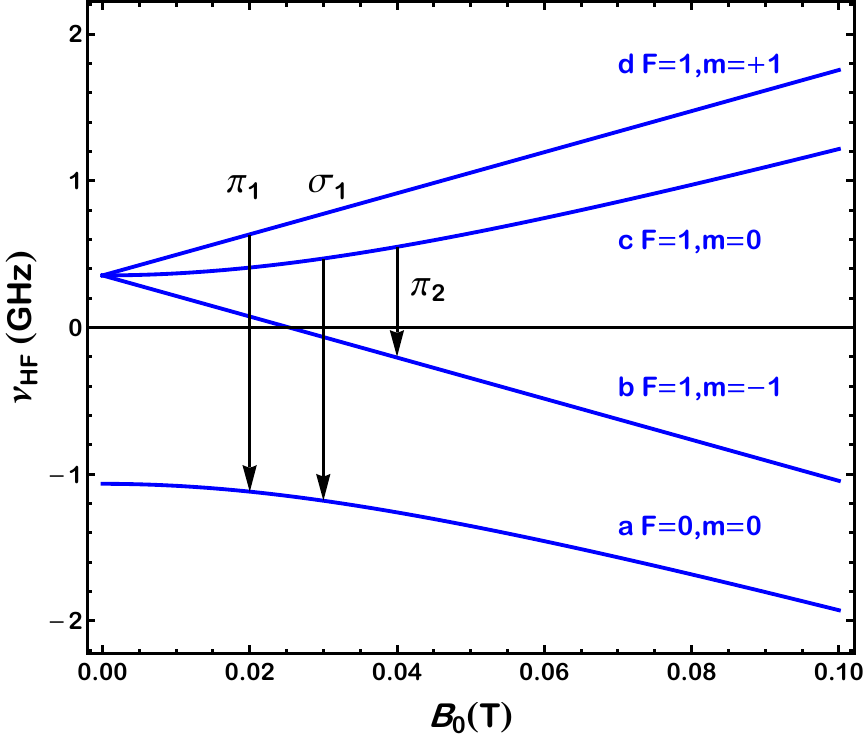}
    \caption{Left: Level diagram of hydrogen and antihydrogen. Right: Breit-Rabi diagram of antihydrogen. For hydrogen the quantum numbers $(F,M_F)=(1,-1)$ and $(1,1)$ are exchanged.}
\label{fig:Hlevels}
\end{center}
\end{figure}

The ALPHA collaboration at CERN has in recent years performed several spectroscopy measurements in antihydrogen trapped in a Ioffe-Pritchard trap (IPT) \cite{Andresen:2010jba}: the observation \cite{Ahmadi:2017a} and determination of the 1S-2S two-photon laser transition (cf. Fig.\ref{fig:Hlevels} (left) for the various transitions) with accuracy of $2\times10^{-12}$ \cite{Ahmadi:2018a}, and the observation of $n=1$ hyperfine transitions \cite{Amole:2012bh} and their determination to $4\times10^{-4}$ precision \cite{Ahmadi:2017}. Measurements were also done on the $n=2$ fine structure \cite{Ahmadi:2020} and the Lyman-$\alpha$ transition \cite{Ahmadi:2018b}, which allows the extraction of the Lamb shift in antihydrogen to a moderate precision of 11 \% \cite{Ahmadi:2020}. A first observation of laser cooling using a Lyman-$\alpha$ laser \cite{Baker:2021} promises to achieve narrower line widths and thus higher precision for the 1S-2S transition.

For the 1S-2S transition, the precision that can be ultimately reached in a trap is difficult to estimate. The best measurement of hydrogen using a cold beam in a field-free region has an accuracy of $4.2\times 10^{-15}$ \cite{Parthey:2011ys}. The measurement of \cite{Ahmadi:2018a} with \Hbar\ was done at a background field of $B\sim 1$T (needed to trap \pbar\ before forming \Hbar), which leads to a shift of $\sim600$ kHz ($\widehat{=}\, 2.5\times 10^{-10}$). This shift has to be corrected using QED calculations that depend on the actual value of $B$ which needs to be measured separately. Furthermore,  the velocity distribution of \Hbar\ atoms having finite temperature causes the atoms to move in regions of varying magnetic field inside the trap, leading to a modification of the line shape that has to be modeled using simulations.  
Achieving higher precision will therefore require one to measure H in the same trap, in which case -- since the line shape depends on the velocity distribution of the atoms -- also the temperature has to be controlled. Alternatively, the atoms can be moved into a field-free region for spectroscopy as described in the next section.

In-trap hyperfine spectroscopy is more difficult as the magnetic fields and field gradients necessary for trapping lead to large Zeeman shifts. One can use combinations of electron spin-flip transitions that yield the zero-field HFS splitting \nuHF\ without knowing the magnetic field, as done in \cite{Amole:2012bh,Ahmadi:2017} which extracted \nuHF\ = \nupone\ -- \nuptwo\ (cf. Fig.\ref{fig:Hlevels} (right)).  A next step in precision could be achieved by using double resonance methods as proposed by ALPHA \cite{Hayden:22}, driving one electron spin flip (e.g. \nupone) and one proton spin flip transition (($F,M_F)=(1,0)\rightarrow(1,-1) $) simultaneously and making use of the fact that proton spin-flip transitions have a magnetic-field insensitive point at $B \sim 0.65$ T \cite{Bluhm:1999vq}.

ASACUSA has proposed an in-beam method for hyperfine spectroscopy \cite{Widmann:04,Malbrunot:2017,Widmann:2019}, and shown that the method works at the ppb level with hydrogen \cite{Diermaier:2017}. Based on this result they estimated that with a few thousand antihydrogen atoms, an accuracy of ppm could be reached. A straight-forward extension of the currently used Rabi method would be to employ a Ramsey method of separated oscillatory fields \cite{Ramsey:1949fk,Ramsey:1950ud}, where a further factor $\sim20$ could be gained. The group of T. H{\"a}nsch at the MPQ Munich, has pioneered this method for optical spectroscopy by applying it to a cryocooled hydrogen beam. They observed  Ramsey fringes on the two-photon 1S-2S transition \cite{BGross_1998}, and improved the resolution of the line to 6 parts in 10$^{-13}$, which was the world record at that time for the resolution of a transition for neutral atoms.

\subsubsection{Fountain experiments}

\begin{wrapfigure}{r}{0.45\textwidth}
\begin{center}
    \includegraphics[width=65mm]{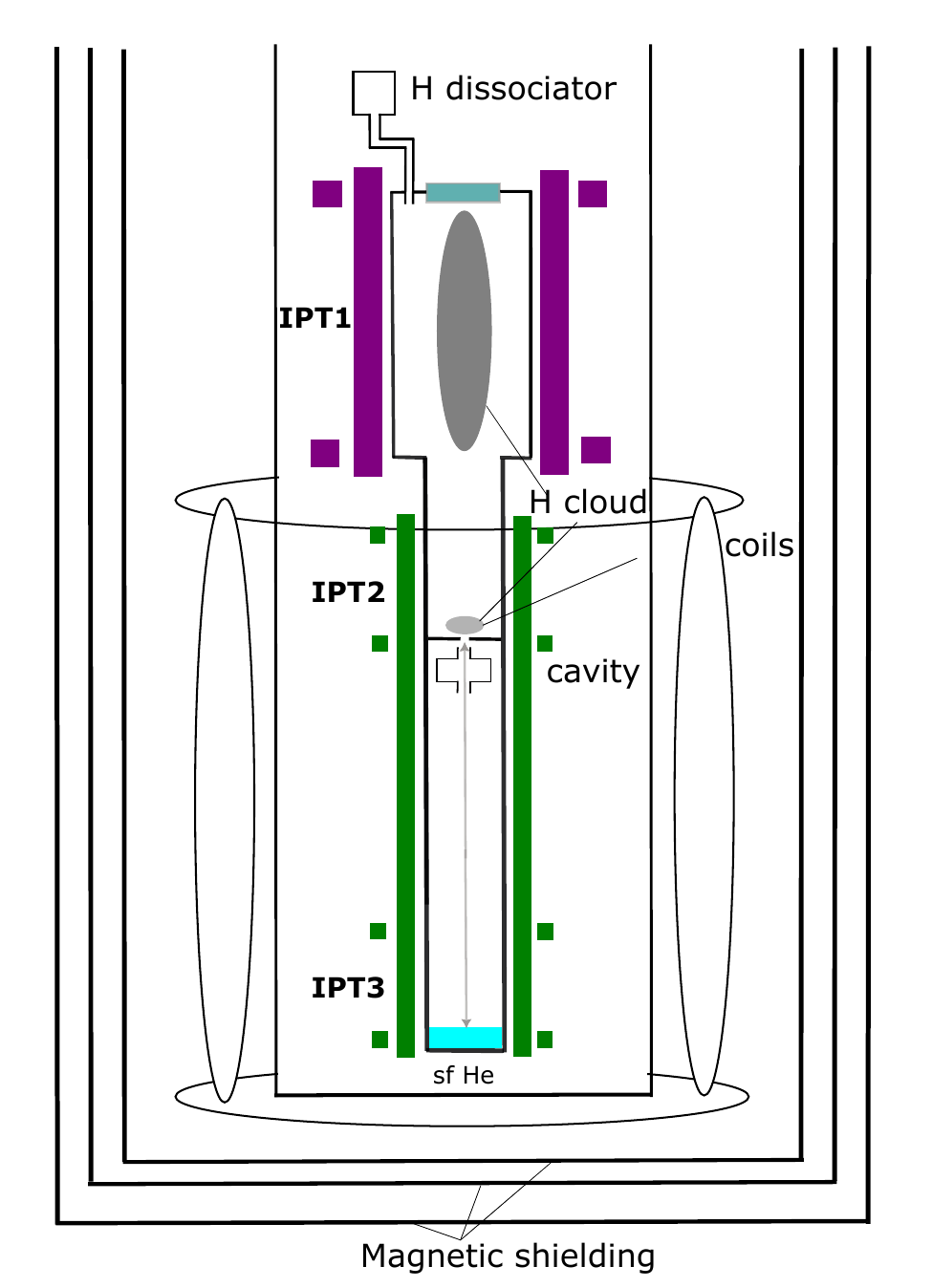}
    \caption{Schematic drawing of a setup to create ultra-cold H through evaporative cooling in a superconducting trap IPT1. H atoms can either be trapped in a normal-conducting trap IPT2 and dropped onto a superfluid He film, or trapped in IPT3 and released to form a fountain. Adopted from \cite{10.1063/5.0070037}.}
\label{fig:ultra-coldH}
\end{center}
\end{wrapfigure}

A further substantial increase in precision can be obtained by moving to much slower atoms in a field free region. The slower the atoms, the longer they interact with the applied laser or microwave radiation and thus the narrower the lines become. This is true as long as the resulting line width is larger than the natural line width of the transition or the line width of the radiation source. The best way towards slower atoms is to follow the path of ultra-cold atoms and to create a fountain from trapped (anti)hydrogen atoms, where (anti)atoms are expelled from the inhomogeneous field region needed for trapping and cooling to a field-free region where spectroscopy can be performed. RF spectroscopy has e.g. been demonstrated in a fountain created from laser-cooled Na atoms \cite{Kasevich:89}, and Cs fountain clocks provide currently the most accurate definition of the SI second (see e.g. \cite{Wynands:2005}). Cs fountain clocks make use of a Ramsey method, where the same cavity is traversed twice, once upwards and twice downwards, reducing systematic effects for a situation with two cavities. The idea of optical Ramsey spectroscopy applied to hydrogen atoms in a fountain is a long term endeavor. Simulations show \cite{Beausoleil86} that a resolution of the natural line width of the 1S-2S transition and thus sub-Hz precision can be obtained.

Indeed, for (anti)hydrogen, the potential energy difference of 1 m in the Earth gravitational field corresponds to a temperature of a few mK, and velocities of a few m/s. Such temperatures can be reached for H in cryogenic traps using evaporative cooling, e.g. as used in the formation of a Bose Einstein Condensate of hydrogen \cite{Fried:98}. As described in sec.~\ref{GravitationalQuantum}~\ref{sec:trappedHspec}, the GRASIAN collaboration \cite{GRASIAN} has built a large octupole IPT and demonstrated its operation \cite{10.1063/5.0070037}. Based on this, experiments can be foreseen that use either trapped H at $T \sim 100\,\mu$K and launch a fountain inside a cryostat, or alternatively let them fall across a cavity onto a superfluid He film where they will be reflected and, when passing the cavity again, perform a Ramsey measurement (cf. Fig.~\ref{fig:ultra-coldH}, more details will be given in sec.~\ref{GravitationalQuantum}~\ref{sec:trappedHspec}). 
% references to sections are incomplete
For hyperfine measurements a well-defined magnetic field environment needs to be created by shielding external fields and applying a homogeneous background field at the position of the atoms.  Such experiments can also be applied to \Hbar\ as proposed by the GBAR collaboration, see sec.~\ref{GravitationalQuantum}~\ref{sec:trappedHspec}. 

HAICU \cite{HAICU} is another recent experiment aiming at creating a fountain interferometer compatible with both hydrogen and antihydrogen atoms. It foresees magnetically compressing (anti)-hydrogen atoms into a mm-scale volume where three-dimensional Doppler-cooling is applied. A next cooling stage is achieved via adiabatic expansion to another trap to reach micro-Kelvin energies before a magnetic launch  into a fountain volume. 

%n atomic fountain can help overcome this limit by interrogating the anti-atoms during free-fall in a field-free volume. A practical antihydrogen fountain requires cooling the anti-atoms to micro-Kelvin energies, such that the volume of a free-falling bunch remains small enough for interrogation (usually via a laser or microwave pulse).

%\subsubsection{Spectroscopy with quantum bounces}

\subsection{Gravitational quantum states}
\label{GravitationalQuantum}

\subsubsection{GQS principle and the historical experiment}

The framework of Gravitational quantum states (GQS) of particles was first derived in 1928 \cite{Breit28} with a massive particle inside a triangular potential well, and soon it became a popular problem for students in many textbooks on quantum mechanics \cite{Goldman60,Landau77}. Only nearly 75 years later, scientists demonstrated this phenomenon through the direct observation of the first GQS on neutron \cite{Nesvizhevsky:2002vv}. Confined on the bottom by the quantum reflections (QR) on a mirror and on top by gravity, Ultra Cold Neutrons (UCNs) were measured to only exist in discrete quantum states, as theoretically expected \cite{Westphal_2007}.  But until now, the GQS of atoms, predicted to exist following the same physics than for the neutrons (n), have never been observed. 
After the observation of GQS of neutrons, active analysis of the peculiarities of these phenomena was initiated and their applications to the search for new physics, such as the searches for extra fundamental short-range interactions \cite{ANTONIADIS2011}, verification of the Weak Equivalence Principle in the quantum regime \cite{OBertolami_2003}, extensions of quantum mechanics \cite{Saha06}, extensions of gravity and space theories or tests of Lorentz invariance 
\cite{Escobar2019b,IVANOV2019134819} were studied. Similar impacts on modern physics is expected to be triggered by the studies of GQS on (anti)-hydrogen, as for instance for the WEP tests \cite{Crepin19_WEP}. The aim of this section is to give a glimpse about some of the possibilities offered by the studies of GQS and QR with anti-atoms/atoms. Compared to neutrons, H atoms (whose mass is similar to n) allow for a larger and slower particle flow allowing a better resolution of these states. They are a good testbench before moving towards the studies of GQS on anti-atoms. Besides, compared to neutrons, the reflection mechanism is different for (anti-)H since it occurs without any contact to the surface, allowing one to probe different interactions \cite{Dufour13}. As (anti)-hydrogen is the simplest atom (only two particles bound together), its properties can be calculated very precisely \cite{Kar05} ; thus it is an ideal probe to test precisely new physics \cite{Buisseret_2007,Kosteleck15}. GQS studies could also be applied to various exotic systems, such as Muonium or Positronium  \cite{GQSPs2015}, and recent development on sources of slow Ps ($\sim$ 10 K) atoms could serve the observation of GQS on these purely leptonic antimatter-matter bound systems \cite{Cooper2016}.

\begin{figure}
\begin{center}
    \includegraphics[width=90mm]{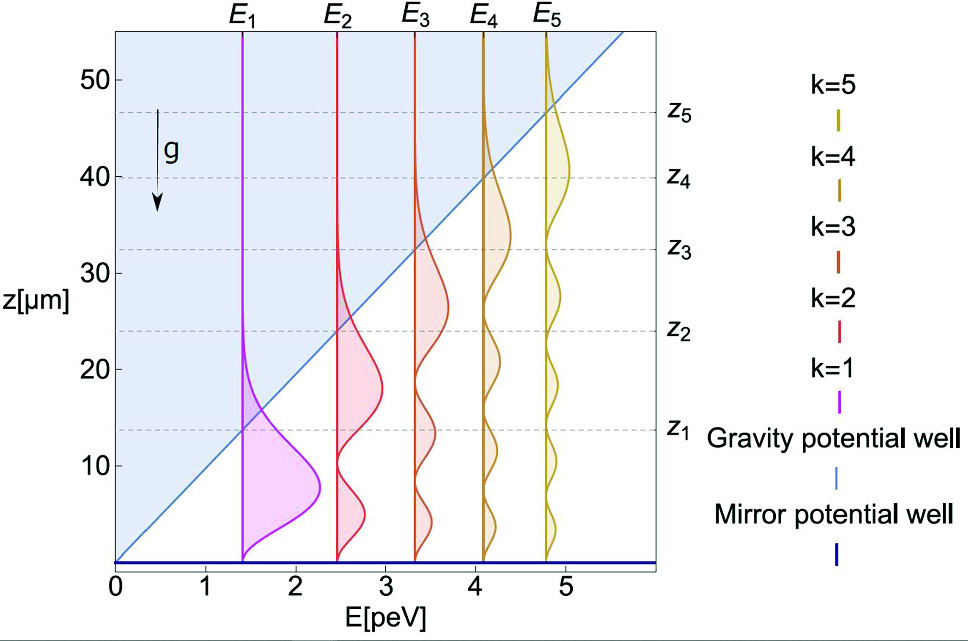}
    \caption{Figure from \cite{Killian23}, cover of EPJD, 77 (March 2023): GQS eigenstates of 1 atomic-mass particle trapped inside the gravity (top) and QR (bottom) potential well.}
\label{Fig1_GQS}
\end{center}
\end{figure}

{\bf GQS principle}: The particle (neutron, H, \Hbar, etc) wavefunction $\psi(z)$ in the Earth’s gravitational field ($V(z)=mgz$) above an ideal mirror is governed by the Schrodinger equation $ \frac{\hbar^2}{2\,m} \frac{d^2 \psi(z)}{dz^2} (E-mgz)\psi(z)=0 $; where $\hbar$ is the reduced Planck constant, $m$ is the particle's mass, $z$ is the height (perfect reflector at $z$=0), $E$ is the energy of the vertical motion of the particle, and $g$ is the acceleration in the Earth’s gravitational field.  The solutions of such equations are given by the Airy functions and the eigenstates are plotted in Fig. \ref{Fig1_GQS}, with typical energy splitting of peV. In short, in order to form GQS, the particles need to be trapped long enough between the gravitational potential and the reflecting surface. Indeed, following the Heisenberg uncertainty principle $\Delta \tau\,\Delta$E $\geq \frac{\hbar}{2}$,  the time of flight $\Delta \tau$ that the atoms spend in the confinement region should be long enough such that the uncertainty on the energy of the states $\Delta$E decreases to be much smaller than the typical energy of the GQS ($\sim$ peV). The typical interaction time is in $\sim $ms. This theoretical statement leads to some stringent experimental requirements (slow enough particles) and explains why the direct observation of QGS remains currently challenging.
Let us point out that GQS of n, H and \Hbar\ are expected to be formed essentially via identical process. First, the only parameter of the particle which enters to the Schrodinger equation for a neutral particle above a perfect mirror is its mass $m$.  Details studies have been performed about the effect of heavier mass particles and how it affects the GQS derivation \cite{BERTOLAMI2006111}, but for our discussion the considered particles n, H and \Hbar\ have nearly the same mass, close to the atomic mass unit.  Second, the  flat surface is nearly a perfect mirror for sufficiently slow H and \Hbar\ due to the high probability of quantum reflection \cite{Voronin11,Dufour13} and the fact that the characteristic range of raising the surface potential, $\approx 10^{-10} -  10^{-9}$~m, is much smaller than the particle wavelength in GQS $\approx 10^{-5} $~m, such that the particles only see an infinite steep potential when approaching the surface, and bounce back due to quantum reflection. %Specific derivations of the potential interaction with each of the mentioned particles have been carefully performed [group Serge].

\subsubsection{Realizing the historical "fly-through" experiment with hydrogen}

\begin{figure}[h]
\centering
\includegraphics[width=120mm]{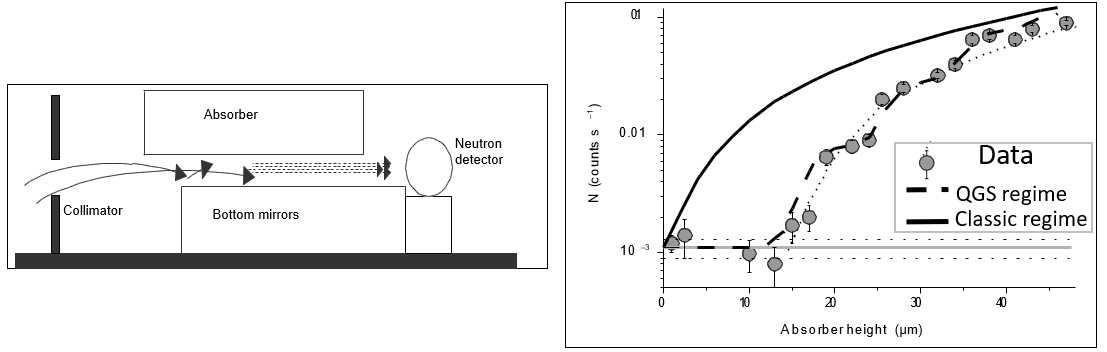}
\caption{(left):  experimental schematic for the detection of GQS of neutrons. (right)  experimental data \cite{Nesvizhevsky:2002vv} compared to the two flux models (classic and quantum regimes).}
\label{Fig2_fly_through}
\end{figure}

The GQS of neutrons have been observed for the first time in 2002 at ILL \cite{Nesvizhevsky:2002vv}. A sketch of the experiment is shown in Fig. \ref{Fig2_fly_through} (left). A flux of collimated slow neutrons impinging at grazing angles to a bottom surface (flat silicon mirror), and on top of it, an “absorber” scatters any neutron that reaches the top surface, such that only the particles with a well-defined quantum state fitting in the gap between the mirror and absorber are detected at the exit. The experiment consists in lifting slowly the absorber above the mirror such that the spacing between the two increases. As shown in Fig. \ref{Fig2_fly_through} (right), in the classical regime, we expect the detected flux to increase continuously as the space between mirror and absorber increases. Whereas, in the quantum regime, the neutrons can only pass through if the gravity quantum states can exist between the mirror and absorber, meaning that the spacing is wider than the typical spread of wavefunction of the GQS. Until reaching the typical spacing of the first QGS, no flux is expected to be detected. Then, as the spacing is large enough, a first flux of neutrons is detected, corresponding to the neutrons into the first GQS ; and in the same way for the next GQS. In case of QGS formation, the flux of particles detected is a step-wise form, as experimentally measured (Fig. \ref{Fig2_fly_through} (right)) for neutrons \cite{Nesvizhevsky:2002vv}.

A new international collaboration, GRASIAN \cite{GRASIAN}, aims to observe for the first time the QGS of hydrogen,  using a fly-through experiment similar to what has been realised with neutrons.  A test-bench experiment is being developed at ETH Zürich in the group of P. Crivelli \cite{Killian23}; results of this experiment will determine the developments  of the rest of the program. % since important parameters for designing the next steps for the collaboration projects can be extracted. 
Mainly, the foreseen goals are to use the GQS to test new physics such as extra fundamental short-range interactions \cite{ANTONIADIS2011} or perform CPT tests \cite{Kosteleck15}, developing similar experiments to antihydrogen (See Sec \ref{secQR} below). Another long-term project would be to trap H \cite{10.1063/5.0070037} for enough time in order to form long-living GQS, and perform precision microwave and optics spectroscopy to resolve the shifts in energy due to the GQS, probing then short-range forces  (see Sec. \ref{sec:trappedHspec} below.).

\subsubsection{Quantum reflection (QR) experiments}
\label{secQR}

Quantum reflection of massive particles is a well-known and theoretically well-described phenomenon \cite{Dufour13}.  Decades ago, hydrogen \cite{Berkhout89,Doyle91,Berkhout93,Yu93} and other systems \cite{Nayak83} have been scattered off liquid helium surfaces while neon \cite{Shimizu01} and sodium atoms \cite{Pasquini04} have been reflected from room temperature surfaces. In several cases the sticking probability was measured and not directly the reflectivity, and in direct measurements at best values of 60$\%$ were observed \cite{Pasquini04}. 
The development of cryogenic sources of H beam opens the possibility of directly measure the high QR ratio theoretically predicted for (anti-) hydrogen, due to rapidly varying Casimir-Polder
potential appearing in the vicinity of the surface and that can lead to specular reflectivities >95~$\%$ \cite{Crepin17,Crepin19_QR,Dufour13,Voronin16}. 
This would allow to ``store'' hydrogen for long times by multiple bounces on surfaces, which in turn would allow spectroscopy of hydrogen with unprecedented accuracy, see Sec. \ref{sec:trappedHspec}. Within the GRASIAN collaboration, an experiment at the Stefan Meyer Institute (Vienna) in the group of E. Widmann, is under development. The idea is to use a cryogenic hydrogen beam ($\sim$ 4 K), collimated and sent at grazing angle onto different surfaces, to first measure the QR coefficient (comparison between specularly reflected and scattered atoms). %The possibility of GQS formation with sufficiently slow atoms undergoing only one QR is under studies [cite paper].

\begin{figure}
\centering
\includegraphics[width=140mm]{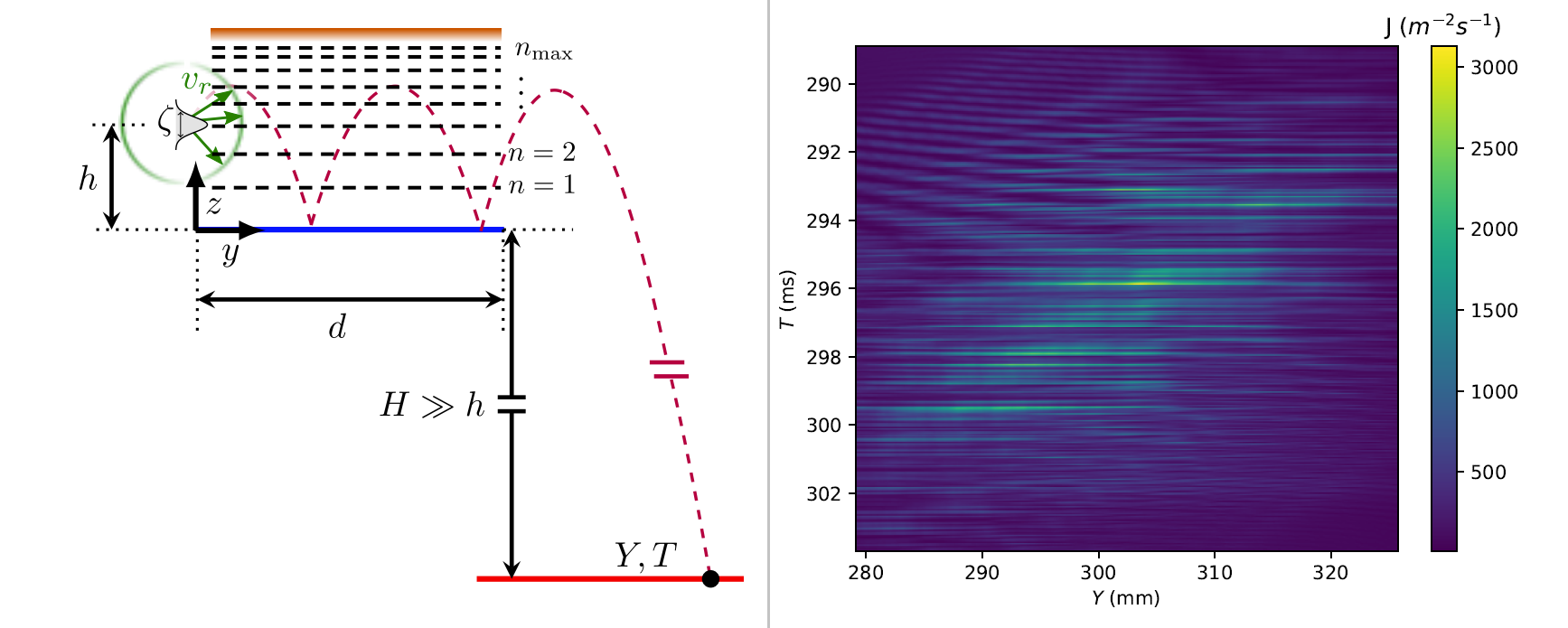}
\caption{Figure  from \cite{Rouselle22}. Left:  schematic representation of the concept of the experiment. The blue horizontal line represents the quantum mirror of radius d. The atoms are released at a mean height h above the mirror with a dispersion $\zeta$ in all space directions. An absorber (orange line) is placed above the mirror allowing nmax Gravitational Quantum States to pass through the device. H is the free-fall height, Y and  T the positions in space and time of the detection events on the detector plate in red. Right: the quantum interferences from the different GQS of the \Hbar\ atoms, lead to the   probability current density J(Y; T) on the detection
plate, for Z = - 30cm and X = 0. Parameters: f$_{RF trap}$ = 20 kHz, $\delta$E = 10 $\mu$eV, $N_{\overline{\mathrm{H}}}=$ 1000.}

\label{Fig3_QR_Gbar}
\end{figure}

QR and GQS formations of ultra-cold anti-atoms find an appealing application for the measurement of the Earth gravitational acceleration, such as for instance, within the GBAR collaboration \cite{Indelicato14}. Shortly, the principle of GBAR is to form an antihydrogen ion (\Hbar$^+$), in order to co-trap it with a laser-cooled berrylium atom, in a Paul trap. Once sympathetically cooled to $\delta E_{Kin}\sim 10\mu eV$ a photodetachment laser shot removes the exceeding positron, leading to the neutral \Hbar\ atom to "fall" under gravity. With a time of flight measurement, the initial proposal aimed for a 1$\%$ resolution on the "g" measurement. Due to the extremely low initial temperature of the \Hbar\ for the GBAR measurement, intensive studies have been performed to understand the QR that will occur on the vacuum chamber's walls of the experiment \cite{Dufour13,Voronin16}.  An interesting proposal is to make use of these highly probable quantum reflections that will enable the formation of GQS after several quantum bounces and improve the direct tests of the Weak Equivalent Principle \cite{PhysRevA.95.032501,Crepin19_WEP}. Indeed, careful calculations have been derived and show that an improvement on the expected sensibility on the measurement on "g" could reach  $10^{-5 } $ \cite{Rouselle22} 
relative uncertainty using only $N_{\overline{\mathrm{H}}}=$ 1000 anti-atoms. To put this number in perspective, as of now, the direct experimental bounds on the ratio of gravitational mass and inertial mass for antihydrogen are : $-65<(\frac{M_g}{M})_{\bar{\textrm{H}}}<110$ \cite{ALPHA_firstg}. The principle of the  measurement is sketched in Fig. \ref{Fig3_QR_Gbar} left, where the initial wavepacket $\zeta$ of one \Hbar\ is shown, that takes into account the photon-recoil from the photo-detachment process as well as the micro-motion from the RF trap (using the frequency trap $f$ as a simulation parameter). The anti-atom falls under the gravitational Earth's field  and undergoes several QR (red-dashed curve) on the bottom surface of the vacuum chamber, leading to the formation of gravitational quantum states (depicted in black dashed energy levels) and to some interferences in the momentum distributions of the anti-atoms. After a free fall of H~$=30$~cm, the interferences can be detected using a position-sensitive detection (annihilation of the \Hbar\ can be positioned with 10 $\mu$m resolution), a typical reconstructed signal is plotted in Fig.\ref{Fig3_QR_Gbar}, right. A new design of the free-fall chamber could enhance even more the resolution of g with fewer \Hbar\ down to N=10 only  \cite{Rouselle22_fewN}.

%\subsubsection{GQS spectroscopy of trapped (anti)-H}
\subsubsection{Spectroscopy with ultra-cold trapped (anti)-H}
\label{sec:trappedHspec}
%\noindent\textbf{EW: title should not be GQS as we do not only talk about making transitions between GQS states}

As mentioned in Sec. \ref{spectrofountain}, the highly resolved spectroscopy of anti-H/H atoms is a key method for CPT and direct tests of the QED predictions. We dealt with the possibility of doing spectroscopic measurement in Ramsey-type fountain based experiments previously. To continue the overview on possible GQS experimental applications, let's discuss here another project that aims to trap and form 100 µK hydrogen atoms.  Trapping and cooling hydrogen atoms at low temperature and high enough density to form BEC was only performed once by the group of D. Kleppner at MIT \cite{Fried:98} and in the two-dimensional H gas at the University of Turku \cite{Safonov98_BEC_Turku} in 1998.

A new proposal of trapping H atoms using a magnetic trap for cooling a large number of H atoms below 1 mK has been recently developed within GRASIAN \cite{Vasiliev19}, in the Turku University in the group of S. Vasiliev. From this preliminary trap, the cold atoms will be transferred to a secondary trap IPT2 for selecting only the lowest velocities 
\cite{10.1063/5.0070037}. Then, the ultra-slow atoms will be released and fall to bounce onto superfluid helium coated surface, being trapped by gravity on top and QR on bottom. A vertical magnetic field will ensure the transverse confinement, forming what has been called a magneto-gravitional trap \cite{MGT_20}. The liquid He surface presents the weakest interaction potential and provides nearly perfect reflecting conditions for low-energy atoms \cite{Crepin17,Crepin19_QR}, allowing for long trapping time > 1s, and enabling the formation of long-living GQS states (high n). Within this magneto-gravitational trap, highly resolved spectroscopy of 1S-2S states of H atoms (natural linewidth of $\sim$ 1.2 Hz or 0.125 s lifetime), will be possible, with drastically reduced second-order Doppler effect, one of the main systematic effects limiting the current measurement \cite{Parthey:2011ys}. %as long as the gradient of B field for the trap will be under-controlled to calculate the Zeeman shifts induced on the transitions. For the long-term project and extremely high resolve measurements, this weak magnetic field can be switched off quickly to perform the spectroscopy. 
The ultimate goal would be to be able to perform the  spectroscopy of the GQS energy splitting ($\sim$ 200 Hz difference between the n=1 and n=2 GQS states on hydrogens), with various spectroscopic methods, such as Ramsey-type spectroscopy described in \cite{MGT_20}. Building upon ALPHA's Ioffe Pritchard trap (magnetic bottle type trap) for antihydrogen atoms, the laser cooling developed in the same apparatus \cite{Baker:2021}, and the 1S-2S spectroscopy methods that are currently being improved to eventually match the precision achieved in H atoms \cite{Ahmadi:2018a}, 
%one could consider the long-term goal 
could open a path for performing the GQS spectroscopic measurements on ultra-cold antihydrogen atoms as well. This would allow an ultimate direct comparison with \Hbar/H predictions for the GQS energy shifts, where one could dream of probing the Casimir-Polder shifts on the quantum levitated states \cite{PhysRevA.95.032501}. Directly measuring the energy shifts induced by the formation of the GQS for both hydrogen and its anti-atom will enable one to test the theory on short-distance forces \cite{Lambrecht11,Crepin19_QR}.

%% EW HFS spectroscopy

Hyperfine spectroscopy of H could also be performed from the ultra-cold hydrogen source described above. Here it will not be possible to use the GMT trap as the strong magnetic field gradients would lead to large frequency shifts. HFS measurements need to be performed in a region of small but extremely homogeneous magnetic field. As indicated in Fig.~\ref{fig:ultra-coldH}, the cold atoms held at the bottom of IPT2 can be released,  traverse a microwave cavity and be reflected from a superfluid He film. On the way up they pass the cavity a second time, creating a Ramsey measurement. The atoms can be retrapped at IPT2, or let fall again to repeat the process. For a fall distance of 10 cm, the QR coefficient is $\sim 0.88$ \cite{Crepin17} and the round-turn time for one bounce is 0.2 s. Thus it should be possible to reach interaction times of order 1 second, promising sub-Hz precision. For lateral confinement a tube covered with superfluid He can be used. 

Theory also predicts lifetimes of H/\Hbar\ in GQS above superfluid He of more than 1 s \cite{Crepin17}, allowing alternatively to build a maser with walls coated with superfluid He. 
Both methods can also be used for \Hbar, provided a source of $\sim 100~\mu$K temperature can be prepared using e.g. the method proposed by GBAR \cite{Indelicato14} or HAICU \cite{HAICU}.

\section{Conclusion}
We have reviewed here some of the new experimental developments addressing the matter-antimatter asymmetry puzzle using atomic and molecular systems. These low energy systems have the advantage of offering potentials for unprecedented precision building upon AMO techniques which are being further developed in the search for CP and CPT symmetry violations. 
Some of the endeavours discussed are table-top like experiments (EDM measurements in cryogenic matrices, hydrogen spectroscopy). Others require accelerators facilities to generate exotic species (radioactive molecules and antihydrogen atoms), and  thus represent complementary experimental approaches to address fundamental particle physics questions. 
In the case of CP violation searches with electron EDM  two major next-generation endeavours are being pursued to surpass gas-phase experiments' sensitivities. The first one consists in using short-lived radionuclides in tailored molecular systems to benefit from extremely large enhancement factors. The second one consists in combining very high densities (as offered by solid samples in previous searches) while retaining key properties for quantum control such as optical pumping techniques, using cryogenic matrices. On the CPT tests side, the realm of high precision spectroscopy with antihydrogen atoms and the successful attempts at laser cooling have widened the horizons and renewed interest in hydrogen fountains and slow atomic beams in general in particular in view of the increased precision and perspective for new physics searches  that quantum gravitational states measurement would offer.
These ambitious new efforts highlight the dynamism of the field and the potentials offered by high-precision small- to mid-scale experiments in the search for physics beyond the Standard Model.

\section*{Acknowledgement}
We acknowledge S.~Lahs, V.~Nesvizhevsky, F.~Nez,  and S. Reynaud  for fruitful discussions. This work was in part supported by the Agence Nationale de la Recherche (ANR) under the project ANR-21-CE30 -0028-01, the Austrian Science Fund (FWF) [W1252-N27] (Doktoratskolleg Particles and Interactions), and the Natural Sciences and Engineering Council of Canada (NSERC). TRIUMF receives federal funding via a contribution agreement with the National Research Council of Canada.

\bibliographystyle{unsrt}
\bibliography{bio}

\begin{thebibliography}{100}

\bibitem{Sakharov:1967dj}
A.~D. Sakharov.
\newblock {Violation of CP Invariance, C asymmetry, and baryon asymmetry of the universe}.
\newblock {\em Pisma Zh. Eksp. Teor. Fiz.}, 5:32--35, 1967.

\bibitem{Agrawal_2021}
P.~Agrawal, M.~Bauer, J.~Beacham, A.~Berlin, A.~Boyarsky, S.~Cebrian, X.~Cid-Vidal, D.~d'Enterria, A.~De Roeck, M.~Drewes, B.~Echenard, M.~Giannotti, G.~F. Giudice, S.~Gninenko, S.~Gori, E.~Goudzovski, J.~Heeck, P.~Hernandez, M.~Hostert, I.~G. Irastorza, A.~Izmaylov, J.~Jaeckel, F.~Kahlhoefer, S.~Knapen, G.~Krnjaic, G.~Lanfranchi, J.~Monroe, V.~I.~Martinez Outschoorn, J.~Lopez-Pavon, S.~Pascoli, M.~Pospelov, D.~Redigolo, A.~Ringwald, O.~Ruchayskiy, J.~Ruderman, H.~Russell, J.~Salfeld-Nebgen, P.~Schuster, M.~Shaposhnikov, L.~Shchutska, J.~Shelton, Y.~Soreq, Y.~Stadnik, J.~Swallow, K.~Tobioka, and Y.-D. Tsai.
\newblock Feebly-interacting particles: {FIPs} 2020 workshop report.
\newblock {\em The European Physical Journal C}, 81(11), nov 2021.

\bibitem{Agarwalla2023}
Sanjib~Kumar Agarwalla, Sudipta Das, Alessio Giarnetti, Davide Meloni, and Masoom Singh.
\newblock Enhancing sensitivity to leptonic cp violation using complementarity among dune, t2hk, and t2hkk.
\newblock {\em The European Physical Journal C}, 83(8):694, 2023.

\bibitem{doi:10.1146/annurev-nucl-102014-021939}
M.V. Diwan, V.~Galymov, X.~Qian, and A.~Rubbia.
\newblock Long-baseline neutrino experiments.
\newblock {\em Annual Review of Nuclear and Particle Science}, 66(1):47--71, 2016.

\bibitem{doi:10.1146/annurev-nucl-101918-023407}
Michelle~J. Dolinski, Alan~W.P. Poon, and Werner Rodejohann.
\newblock Neutrinoless double-beta decay: Status and prospects.
\newblock {\em Annual Review of Nuclear and Particle Science}, 69(1):219--251, 2019.

\bibitem{safronova2018search}
MS~Safronova, D~Budker, D~DeMille, Derek F~Jackson Kimball, A~Derevianko, and Charles~W Clark.
\newblock Search for new physics with atoms and molecules.
\newblock {\em Reviews of Modern Physics}, 90(2):025008, 2018.

\bibitem{alarcon2022electric}
Ricardo Alarcon, Jim Alexander, Vassilis Anastassopoulos, Takatoshi Aoki, Rick Baartman, Stefan Bae{\ss}ler, Larry Bartoszek, Douglas~H Beck, Franco Bedeschi, Robert Berger, et~al.
\newblock Electric dipole moments and the search for new physics.
\newblock {\em arXiv preprint arXiv:2203.08103}, 2022.

\bibitem{RevModPhys.53.1}
A.~D. Dolgov and Ya.~B. Zeldovich.
\newblock Cosmology and elementary particles.
\newblock {\em Rev. Mod. Phys.}, 53:1--41, Jan 1981.

\bibitem{COHEN1988913}
Andrew~G. Cohen and David~B. Kaplan.
\newblock Spontaneous baryogenesis.
\newblock {\em Nuclear Physics B}, 308(4):913--928, 1988.

\bibitem{Alan_Kosteleck__1996}
V.~Alan Kosteleck{\'{y}} and R.~Potting.
\newblock Expectation values, lorentz invariance, and {CPT} in the open bosonic string.
\newblock {\em Physics Letters B}, 381(1-3):89--96, jul 1996.

\bibitem{Ellis_2013}
John Ellis, Nick~E. Mavromatos, and Sarben Sarkar.
\newblock Environmental {CPT} violation in an expanding universe in string theory.
\newblock {\em Physics Letters B}, 725(4-5):407--411, oct 2013.

\bibitem{PhysRevD.55.6760}
Don Colladay and V.~Alan Kosteleck\'y.
\newblock $\mathrm{CPT}$ violation and the standard model.
\newblock {\em Phys. Rev. D}, 55:6760--6774, Jun 1997.

\bibitem{Ahmadi:2018a}
M.~Ahmadi, B.~X.~R. Alves, C.~J. Baker, W.~Bertsche, A.~Capra, C.~Carruth, C.~L. Cesar, M.~Charlton, S.~Cohen, R.~Collister, S.~Eriksson, A.~Evans, N.~Evetts, J.~Fajans, T.~Friesen, M.~C. Fujiwara, D.~R. Gill, J.~S. Hangst, W.~N. Hardy, M.~E. Hayden, C.~A. Isaac, M.~A. Johnson, J.~M. Jones, S.~A. Jones, S.~Jonsell, A.~Khramov, P.~Knapp, L.~Kurchaninov, N.~Madsen, D.~Maxwell, J.~T.~K. McKenna, S.~Menary, T.~Momose, J.~J. Munich, K.~Olchanski, A.~Olin, P.~Pusa, C.~O. Rasmussen, F.~Robicheaux, R.~L. Sacramento, M.~Sameed, E.~Sarid, D.~M. Silveira, G.~Stutter, C.~So, T.~D. Tharp, R.~I. Thompson, D.~P. van~der Werf, and J.~S. Wurtele.
\newblock Characterization of the {1S}–{2S} transition in antihydrogen.
\newblock {\em Nature}, 557(7703):71--75, May 2018.

\bibitem{Baker:2021}
C.~J. Baker, W.~Bertsche, A.~Capra, C.~Carruth, C.~L. Cesar, M.~Charlton, A.~Christensen, R.~Collister, A.~Cridland Mathad, S.~Eriksson, A.~Evans, N.~Evetts, J.~Fajans, T.~Friesen, M.~C. Fujiwara, D.~R. Gill, P.~Grandemange, P.~Granum, J.~S. Hangst, W.~N. Hardy, M.~E. Hayden, D.~Hodgkinson, E.~Hunter, C.~A. Isaac, M.~A. Johnson, J.~M. Jones, S.~A. Jones, S.~Jonsell, A.~Khramov, P.~Knapp, L.~Kurchaninov, N.~Madsen, D.~Maxwell, J.~T.~K. McKenna, S.~Menary, J.~M. Michan, T.~Momose, P.~S. Mullan, J.~J. Munich, K.~Olchanski, A.~Olin, J.~Peszka, A.~Powell, P.~Pusa, C.{\O}. Rasmussen, F.~Robicheaux, R.~L. Sacramento, M.~Sameed, E.~Sarid, D.~M. Silveira, D.~M. Starko, C.~So, G.~Stutter, T.~D. Tharp, A.~Thibeault, R.~I. Thompson, D.~P. van~der Werf, and J.~S. Wurtele.
\newblock Laser cooling of antihydrogen atoms.
\newblock {\em Nature}, 592(7852):35--42, April 2021.

\bibitem{blum2022fundamental}
T~Blum, P~Winter, T~Bhattacharya, TY~Chen, V~Cirigliano, D~DeMille, A~Gerarci, NR~Hutzler, TM~Ito, O~Kim, et~al.
\newblock Fundamental physics in small experiments.
\newblock {\em arXiv preprint arXiv:2209.08041}, 2022.

\bibitem{boeschoten2023perspectives}
Alexander Boeschoten and Lorenz Willmann.
\newblock Perspectives on electric dipole moments of atoms and molecules.
\newblock In {\em EPJ Web of Conferences}, volume 282, page 01019. EDP Sciences, 2023.

\bibitem{Fukuyama2012}
Takeshi Fukuyama.
\newblock Searching for new physics beyond the standard model in electric dipole moment.
\newblock {\em International Journal of Modern Physics A}, 27(16):1230015, June 2012.

\bibitem{ACME2018}
V.~Andreev, D.~G. Ang, D.~DeMille, J.~M. Doyle, G.~Gabrielse, J.~Haefner, N.~R. Hutzler, Z.~Lasner, C.~Meisenhelder, B.~R. O'Leary, C.~D. Panda, A.~D. West, E.~P. West, X.~Wu, and ACME Collaboration.
\newblock {\em Nature}, 562(7727):355--360, 2018.

\bibitem{HfF_eEDM2023}
Tanya~S. Roussy, Luke Caldwell, Trevor Wright, William~B. Cairncross, Yuval Shagam, Kia~Boon Ng, Noah Schlossberger, Sun~Yool Park, Anzhou Wang, Jun Ye, and Eric~A. Cornell.
\newblock An improved bound on the electron’s electric dipole moment.
\newblock {\em Science}, 381(6653):46--50, 2023.

\bibitem{PhysRevA.108.012804}
Luke Caldwell, Tanya~S. Roussy, Trevor Wright, William~B. Cairncross, Yuval Shagam, Kia~Boon Ng, Noah Schlossberger, Sun~Yool Park, Anzhou Wang, Jun Ye, and Eric~A. Cornell.
\newblock Systematic and statistical uncertainty evaluation of the ${\mathrm{hff}}^{+}$ electron electric dipole moment experiment.
\newblock {\em Phys. Rev. A}, 108:012804, Jul 2023.

\bibitem{PhysRevA.105.022823}
Kia~Boon Ng, Yan Zhou, Lan Cheng, Noah Schlossberger, Sun~Yool Park, Tanya~S. Roussy, Luke Caldwell, Yuval Shagam, Antonio~J. Vigil, Eric~A. Cornell, and Jun Ye.
\newblock Spectroscopy on the electron-electric-dipole-moment--sensitive states of ${\mathrm{thf}}^{+}$.
\newblock {\em Phys. Rev. A}, 105:022823, Feb 2022.

\bibitem{RadMoleculesWhitePaper2023}
Gordon Arrowsmith-Kron and et~al.
\newblock Opportunities for fundamental physics research with radioactive molecules.
\newblock {\em arXiv:2302.02165}, 2023.

\bibitem{PhysRevC.91.035502}
Timothy Chupp and Michael Ramsey-Musolf.
\newblock Electric dipole moments: A global analysis.
\newblock {\em Phys. Rev. C}, 91:035502, Mar 2015.

\bibitem{JBehr_octupoleDeformed2022}
J.A. Behr.
\newblock Nuclei with enhanced schiff moments in practical elements for atomic and molecular edm measurements.
\newblock {\em arXiv:2203.06758}, 2022.

\bibitem{PhysRev.132.2194}
L.~I. Schiff.
\newblock Measurability of nuclear electric dipole moments.
\newblock {\em Phys. Rev.}, 132:2194--2200, Dec 1963.

\bibitem{PhysRevLett.76.4316}
N.~Auerbach, V.~V. Flambaum, and V.~Spevak.
\newblock Collective t- and p-odd electromagnetic moments in nuclei with octupole deformations.
\newblock {\em Phys. Rev. Lett.}, 76:4316--4319, Jun 1996.

\bibitem{PhysRevLett.116.161601}
B.~Graner, Y.~Chen, E.~G. Lindahl, and B.~R. Heckel.
\newblock {\em Phys. Rev. Lett.}, 116:161601, 2016.

\bibitem{PhysRevLett.114.233002}
R.~H. Parker, M.~R. Dietrich, M.~R. Kalita, N.~D. Lemke, K.~G. Bailey, M.~Bishof, J.~P. Greene, R.~J. Holt, W.~Korsch, Z.-T. Lu, P.~Mueller, T.~P. O'Connor, and J.~T. Singh.
\newblock {\em Phys. Rev. Lett.}, 114:233002, 2015.

\bibitem{Blaum_2013}
Klaus Blaum, Jens Dilling, and Wilfried Nörtershäuser.
\newblock Precision atomic physics techniques for nuclear physics with radioactive beams.
\newblock {\em Physica Scripta}, 2013(T152):014017, jan 2013.

\bibitem{PhysRevLett.101.202501}
M.~Smith, M.~Brodeur, T.~Brunner, S.~Ettenauer, A~Lapierre, R.~Ringle, V.~L. Ryjkov, F.~Ames, P.~Bricault, G.~W.~F. Drake, P.~Delheij, D.~Lunney, F.~Sarazin, and J.~Dilling.
\newblock First penning-trap mass measurement of the exotic halo nucleus $^{11}\mathrm{Li}$.
\newblock {\em Phys. Rev. Lett.}, 101:202501, Nov 2008.

\bibitem{doi:10.1126/science.1225636}
E.~Minaya Ramirez, D.~Ackermann, K.~Blaum, M.~Block, C.~Droese, Ch.~E. Düllmann, M.~Dworschak, M.~Eibach, S.~Eliseev, E.~Haettner, F.~Herfurth, F.~P. Heßberger, S.~Hofmann, J.~Ketelaer, G.~Marx, M.~Mazzocco, D.~Nesterenko, Yu.~N. Novikov, W.~R. Plaß, D.~Rodríguez, C.~Scheidenberger, L.~Schweikhard, P.~G. Thirolf, and C.~Weber.
\newblock Direct mapping of nuclear shell effects in the heaviest elements.
\newblock {\em Science}, 337(6099):1207--1210, 2012.

\bibitem{YANG2023104005}
X.F. Yang, S.J. Wang, S.G. Wilkins, and R.F.~Garcia Ruiz.
\newblock Laser spectroscopy for the study of exotic nuclei.
\newblock {\em Progress in Particle and Nuclear Physics}, 129:104005, 2023.

\bibitem{PhysRevLett.126.023002}
M.~Fan, C.~A. Holliman, X.~Shi, H.~Zhang, M.~W. Straus, X.~Li, S.~W. Buechele, and A.~M. Jayich.
\newblock {\em Phys. Rev. Lett.}, 126:023002, 2021.

\bibitem{GarciaRuiz2020}
R.~F. Garcia~Ruiz, R.~Berger, J.~Billowes, C.~L. Binnersley, M.~L. Bissell, A.~A. Breier, A.~J. Brinson, K.~Chrysalidis, T.~E. Cocolios, B.~S. Cooper, K.~T. Flanagan, T.~F. Giesen, R.~P. de~Groote, S.~Franchoo, F.~P. Gustafsson, T.~A. Isaev, {\'A}.~Koszor{\'u}s, G.~Neyens, H.~A. Perrett, C.~M. Ricketts, S.~Rothe, L.~Schweikhard, A.~R. Vernon, K.~D.~A. Wendt, F.~Wienholtz, S.~G. Wilkins, and X.~F. Yang.
\newblock {\em Nature}, 581(7809):396--400, 2020.

\bibitem{PhysRevLett.127.033001}
S.~M. Udrescu, A.~J. Brinson, R.~F.~Garcia Ruiz, K.~Gaul, R.~Berger, J.~Billowes, C.~L. Binnersley, M.~L. Bissell, A.~A. Breier, K.~Chrysalidis, T.~E. Cocolios, B.~S. Cooper, K.~T. Flanagan, T.~F. Giesen, R.~P. de~Groote, S.~Franchoo, F.~P. Gustafsson, T.~A. Isaev, \'A. Koszor\'us, G.~Neyens, H.~A. Perrett, C.~M. Ricketts, S.~Rothe, A.~R. Vernon, K.~D.~A. Wendt, F.~Wienholtz, S.~G. Wilkins, and X.~F. Yang.
\newblock {\em Phys. Rev. Lett.}, 127:033001, 2021.

\bibitem{PhysRevA.82.052521}
T.~A. Isaev, S.~Hoekstra, and R.~Berger.
\newblock {\em Phys. Rev. A}, 82:052521, 2010.

\bibitem{PhysRevA.90.052513}
A.~D. Kudashov, A.~N. Petrov, L.~V. Skripnikov, N.~S. Mosyagin, T.~A. Isaev, R.~Berger, and A.~V. Titov.
\newblock {\em Phys. Rev. A}, 90:052513, 2014.

\bibitem{Isaev_2017}
T~A Isaev et~al.
\newblock {\em J. Phys. B: At. Mol. Opt. Phys}, 50(22):225101, 2017.

\bibitem{PhysRevLett.119.133002}
Ivan Kozyryev and Nicholas~R. Hutzler.
\newblock {\em Phys. Rev. Lett.}, 119:133002, 2017.

\bibitem{PhysRevA.77.024501}
V.~V. Flambaum.
\newblock {\em Phys. Rev. A}, 77:024501, 2008.

\bibitem{PhysRevA.99.052502}
N.~M. Fazil, V.~S. Prasannaa, K.~V.~P. Latha, M.~Abe, and B.~P. Das.
\newblock {\em Phys. Rev. A}, 99:052502, 2019.

\bibitem{PhysRevLett.126.023003}
Phelan Yu and Nicholas~R. Hutzler.
\newblock {\em Phys. Rev. Lett.}, 126:023003, 2021.

\bibitem{10.1063/5.0159888}
Leonid~V. Skripnikov, Alexander~V. Oleynichenko, Andréi Zaitsevskii, Nikolai~S. Mosyagin, Michail Athanasakis-Kaklamanakis, Mia Au, and Gerda Neyens.
\newblock {Ab initio study of electronic states and radiative properties of the AcF molecule}.
\newblock {\em The Journal of Chemical Physics}, 159(12):124301, 09 2023.

\bibitem{PhysRevC.99.035501}
V.~V. Flambaum.
\newblock {\em Phys. Rev. C}, 99:035501, 2019.

\bibitem{PhysRevA.101.042504}
V.~V. Flambaum and V.~A. Dzuba.
\newblock {\em Phys. Rev. A}, 101:042504, 2020.

\bibitem{AU2023375}
M.~Au, M.~Athanasakis-Kaklamanakis, L.~Nies, J.~Ballof, R.~Berger, K.~Chrysalidis, P.~Fischer, R.~Heinke, J.~Johnson, U.~Köster, D.~Leimbach, B.~Marsh, M.~Mougeot, B.~Reich, J.~Reilly, E.~Reis, M.~Schlaich, Ch. Schweiger, L.~Schweikhard, S.~Stegemann, J.~Wessolek, F.~Wienholtz, S.G. Wilkins, W.~Wojtaczka, Ch.E. Düllmann, and S.~Rothe.
\newblock In-source and in-trap formation of molecular ions in the actinide mass range at cern-isolde.
\newblock {\em Nuclear Instruments and Methods in Physics Research Section B: Beam Interactions with Materials and Atoms}, 541:375--379, 2023.

\bibitem{PhysRevResearch.4.033229}
S.~Sels, F.~M. Maier, M.~Au, P.~Fischer, C.~Kanitz, V.~Lagaki, S.~Lechner, E.~Leistenschneider, D.~Leimbach, E.~M. Lykiardopoulou, A.~A. Kwiatkowski, T.~Manovitz, Y.~N. Vila~Gracia, G.~Neyens, P.~Plattner, S.~Rothe, L.~Schweikhard, M.~Vilen, R.~N. Wolf, and S.~Malbrunot-Ettenauer.
\newblock Doppler and sympathetic cooling for the investigation of short-lived radioactive ions.
\newblock {\em Phys. Rev. Res.}, 4:033229, Sep 2022.

\bibitem{Klos_2022}
Jacek Kłos, Hui Li, Eite Tiesinga, and Svetlana Kotochigova.
\newblock Prospects for assembling ultracold radioactive molecules from laser-cooled atoms.
\newblock {\em New Journal of Physics}, 24(2):025005, feb 2022.

\bibitem{battard2023cesium}
Thomas Battard, Sebastian Lahs, Claudine Cr{\'e}pin, and Daniel Comparat.
\newblock Cesium atoms in cryogenic argon matrix.
\newblock {\em arXiv preprint arXiv:2305.11947}, 2023.

\bibitem{lambo2023calculation}
Ricardo~L Lambo, Gregory~K Koyanagi, Anita Ragyanszki, Marko Horbatsch, Rene Fournier, and Eric~A Hessels.
\newblock Calculation of the local environment of a barium monofluoride molecule in an argon matrix: a step towards using matrix-isolated baf for determining the electron electric dipole moment.
\newblock {\em Molecular Physics}, 121(6):e2198044, 2023.

\bibitem{browne1961effects}
ME~Browne.
\newblock Effects of applied electric fields on paramagnetic resonance in chrome alums.
\newblock {\em Phys. Rev.}, 121(6):1699, 1961.

\bibitem{royce1963linear}
EB~Royce and N~Bloembergen.
\newblock Linear electric shifts in the paramagnetic resonance of al 2 o 3: Cr and mgo: Cr.
\newblock {\em Phy. Rev.}, 131(5):1912, 1963.

\bibitem{vasil1978measurement}
BV~Vasil'Ev and EV~Kolycheva.
\newblock Measurement of the electric dipole moment of the electron with a quantum interferometer.
\newblock {\em Sov. Phys.-JETP (Engl. Transl.);(United States)}, 47(2), 1978.

\bibitem{heidenreich2005limit}
BJ~Heidenreich, OT~Elliott, ND~Charney, KA~Virgien, AW~Bridges, MA~McKeon, SK~Peck, D~Krause~Jr, JE~Gordon, LR~Hunter, et~al.
\newblock Limit on the electron electric dipole moment in gadolinium-iron garnet.
\newblock {\em Phys. Rev. Lett.}, 95(25):253004, 2005.

\bibitem{kim2015new}
YJ~Kim, C-Y Liu, SK~Lamoreaux, G~Visser, B~Kunkler, AN~Matlashov, JC~Long, and TG~Reddy.
\newblock New experimental limit on the electric dipole moment of the electron in a paramagnetic insulator.
\newblock {\em Phys. Rev. D}, 91(10):102004, 2015.

\bibitem{eckel2012limit}
S~Eckel, AO~Sushkov, and SK~Lamoreaux.
\newblock Limit on the electron electric dipole moment using paramagnetic ferroelectric eu 0.5 ba 0.5 tio 3.
\newblock {\em Phys. Rev. Lett.}, 109(19):193003, 2012.

\bibitem{pryor1987artificial}
Craig Pryor and Frank Wilczek.
\newblock “artificial vacuum” for t-violation experiment.
\newblock {\em Physics Letters B}, 194(1):137--140, 1987.

\bibitem{hinds1991testing}
EA~Hinds and K~Sangster.
\newblock Testing time-reversal symmetry with molecules.
\newblock In {\em AIP Conference Proceedings}, volume 270, pages 77--83. American Institute of Physics, 1991.

\bibitem{kozlov2006proposal}
MG~Kozlov and Andrei Derevianko.
\newblock Proposal for a sensitive search for the electric dipole moment of the electron with matrix-isolated radicals.
\newblock {\em Physical review letters}, 97(6):063001, 2006.

\bibitem{moroshkin2008atomic}
Peter Moroshkin, Adrian Hofer, and Antoine Weis.
\newblock Atomic and molecular defects in solid 4he.
\newblock {\em Phys. Rep.}, 469(1):1--57, 2008.

\bibitem{ulzega2006theoretical}
Simone Ulzega.
\newblock {\em Theoretical and experimental study of the Stark effect in the ground state of alkali atoms in helium crystals}.
\newblock PhD thesis, Universit{\'e} de Fribourg, 2006.

\bibitem{BarnesEtAl1981}
A.~J. Barnes, W.~J. Orville-Thomas, A.~Müller, and R.~Gaufrès, editors.
\newblock {\em Matrix {Isolation} {Spectroscopy}}.
\newblock Springer Netherlands, Dordrecht, 1981.

\bibitem{BondybeyEtAl1996}
Vladimir~E. Bondybey, Alice~M. Smith, and Jürgen Agreiter.
\newblock New {Developments} in {Matrix} {Isolation} {Spectroscopy}.
\newblock {\em Chemical Reviews}, 96(6):2113--2134, January 1996.
\newblock Publisher: American Chemical Society.

\bibitem{Crepin-GilbertTramer1999}
C.~Crepin-Gilbert and A.~Tramer.
\newblock Photophysics of metal atoms in rare-gas complexes, clusters and matrices.
\newblock {\em International Reviews in Physical Chemistry}, 18(4):485--556, October 1999.

\bibitem{kanagin2013optical}
Andrew~N Kanagin, Sameer~K Regmi, Pawan Pathak, and Jonathan~D Weinstein.
\newblock Optical pumping of rubidium atoms frozen in solid argon.
\newblock {\em Physical Review A}, 88(6):063404, 2013.

\bibitem{upadhyay2019spin}
Sunil Upadhyay, Ugne Dargyte, Vsevolod~D Dergachev, Robert~P Prater, Sergey~A Varganov, Timur~V Tscherbul, David Patterson, and Jonathan~D Weinstein.
\newblock Spin coherence and optical properties of alkali-metal atoms in solid parahydrogen.
\newblock {\em Physical Review A}, 100(6):063419, 2019.

\bibitem{upadhyay2020ultralong}
Sunil Upadhyay, Ugne Dargyte, David Patterson, and Jonathan~D Weinstein.
\newblock Ultralong spin-coherence times for rubidium atoms in solid parahydrogen via dynamical decoupling.
\newblock {\em Physical Review Letters}, 125(4):043601, 2020.

\bibitem{dargyte2021optical}
Ugne Dargyte, David~M Lancaster, and Jonathan~D Weinstein.
\newblock Optical and spin-coherence properties of rubidium atoms trapped in solid neon.
\newblock {\em Physical Review A}, 104(3):032611, 2021.

\bibitem{vutha2018oriented}
AC~Vutha, M~Horbatsch, and EA~Hessels.
\newblock Oriented polar molecules in a solid inert-gas matrix: a proposed method for measuring the electric dipole moment of the electron.
\newblock {\em Atoms}, 6(1):3, 2018.

\bibitem{lambo2021high}
R~Lambo, C-Y Xu, ST~Pratt, H~Xu, JC~Zappala, KG~Bailey, Z-T Lu, P~Mueller, TP~O'Connor, BB~Kamorzin, et~al.
\newblock High-resolution spectroscopy of neutral yb atoms in a solid ne matrix.
\newblock {\em Physical Review A}, 104(6):062809, 2021.

\bibitem{braggio2022spectroscopy}
Caterina Braggio, Roberto Calabrese, Giovanni Carugno, Giuseppe Fiscelli, Marco Guarise, Alen Khanbekyan, Antonio Noto, Roberto Passante, Lucia Rizzuto, Giuseppe Ruoso, et~al.
\newblock Spectroscopy of alkali atoms in solid matrices of rare gases: Experimental results and theoretical analysis.
\newblock {\em Applied Sciences}, 12(13):6492, 2022.

\bibitem{gaire2022excitation}
Vinod Gaire, Colin Parker, Chandra Raman, Jianqiao Li, and Yiting Pei.
\newblock Excitation of magnetic dipole transition of thulium atoms trapped in rare gas crystals.
\newblock In {\em APS Division of Atomic, Molecular and Optical Physics Meeting Abstracts}, volume 2022, pages H04--009, 2022.

\bibitem{li2023optical}
Samuel Li, Harish Ramachandran, Rhys Anderson, and Amar Vutha.
\newblock Optical control of baf molecules trapped in neon ice.
\newblock {\em New Journal of Physics}, 2023.

\bibitem{lambo2023calculationNe}
RL~Lambo, GK~Koyanagi, M~Horbatsch, R~Fournier, and EA~Hessels.
\newblock Calculation of the local environment of a barium monofluoride molecule in a neon matrix.
\newblock {\em arXiv preprint arXiv:2305.10667}, 2023.

\bibitem{ballof2023progress}
J~Ballof, N~Nusgart, P~Lalain, M~Au, R~Heinke, D~Leimbach, S~Stegemann, M~Sch{\"u}tt, S~Rothe, and Jaideep~T Singh.
\newblock Progress towards the frib-edm3-frontend: A tool to provide radioactive molecules from isotope harvesting for fundamental symmetry studies.
\newblock {\em Nuclear Instruments and Methods in Physics Research Section B: Beam Interactions with Materials and Atoms}, 541:224--227, 2023.

\bibitem{azevedo2023platform}
Levi~OA Azevedo, Rodolfo~JS Costa, Wania Wolff, Alvaro~N Oliveira, Rodrigo~L Sacramento, Daniel~M Silveira, and Claudio~L Cesar.
\newblock A platform for trapped cryogenic electrons, anions and cations for fundamental physics and chemical studies.
\newblock {\em arXiv preprint arXiv:2301.13248}, 2023.

\bibitem{anderegg2023quantum}
Lo{\"\i}c Anderegg, Nathaniel~B Vilas, Christian Hallas, Paige Robichaud, Arian Jadbabaie, John~M Doyle, and Nicholas~R Hutzler.
\newblock Quantum control of trapped polyatomic molecules for eedm searches.
\newblock {\em arXiv preprint arXiv:2301.08656}, 2023.

\bibitem{SwansonEtAl1986}
Basil~I Swanson, Llewellyn~H Jones, Scott~A Ekberg, and Herbert~A Fry.
\newblock The use of single crystal rare gas lattices to study homogeneous lineshapes of impurities: Sf$_6$ in xenon.
\newblock {\em Chemical Physics Letters}, 126(5):6, 1986.

\bibitem{guarise2020feasibility}
M~Guarise, C~Braggio, R~Calabrese, G~Carugno, A~Dainelli, A~Khanbekyan, E~Luppi, E~Mariotti, and L~Tomassetti.
\newblock A feasibility study for a low energy threshold particle detector in a xenon crystal.
\newblock {\em Journal of Instrumentation}, 15(03):C03004, 2020.

\bibitem{bhandari2021high}
Ashok Bhandari, Alexandar~P Rollings, Levi Ratto, and Jonathan~D Weinstein.
\newblock High-purity solid parahydrogen.
\newblock {\em Review of Scientific Instruments}, 92(7), 2021.

\bibitem{nexo2019imaging}
nEXO Collaboration.
\newblock Imaging individual barium atoms in solid xenon for barium tagging in nexo.
\newblock {\em Nature}, 569(7755):203--207, 2019.

\bibitem{guarise2020particle}
M~Guarise, C~Braggio, R~Calabrese, G~Carugno, A~Dainelli, A~Khanbekyan, E~Luppi, E~Mariotti, and L~Tomassetti.
\newblock Particle detection in rare gas solids: Demiurgos experiment.
\newblock {\em Nuclear Instruments and Methods in Physics Research Section A: Accelerators, Spectrometers, Detectors and Associated Equipment}, 958:162434, 2020.

\bibitem{guarise2022particle}
Marco Guarise.
\newblock Particle detection in rare gas solid crystals: a feasibility experimental study—exploring new ways for dark matter searches: Exploring new ways for dark matter searches.
\newblock {\em The European Physical Journal Plus}, 137(6):673, 2022.

\bibitem{budker2022quantum}
Dmitry Budker, Thomas Cecil, Timothy~E Chupp, Andrew~A Geraci, Derek F~Jackson Kimball, Shimon Kolkowitz, Surjeet Rajendran, Jaideep~T Singh, and Alexander~O Sushkov.
\newblock Quantum sensors for high precision measurements of spin-dependent interactions.
\newblock {\em arXiv preprint arXiv:2203.09488}, 2022.

\bibitem{kimball2023probing}
Derek F~Jackson Kimball, Dmitry Budker, Timothy~E Chupp, Andrew~A Geraci, Shimon Kolkowitz, Jaideep~T Singh, and Alexander~O Sushkov.
\newblock Probing fundamental physics with spin-based quantum sensors.
\newblock {\em Physical Review A}, 108(1):010101, 2023.

\bibitem{jackson2023search}
Derek~F Jackson~Kimball and Karl Van~Bibber.
\newblock {\em The search for ultralight bosonic dark matter}.
\newblock Springer Nature, 2023.

\bibitem{sikivie2014axion}
P~Sikivie.
\newblock Axion dark matter detection using atomic transitions.
\newblock {\em Physical review letters}, 113(20):201301, 2014.

\bibitem{braggioaxion}
C~Braggio, R~Calabrese, G~Carugno, G~Fiscelli, M~Guarise, A~Khanbekyan, A~Noto, R~Passante, L~Rizzuto, G~Ruoso, et~al.
\newblock Axion detection setup exploiting magnetic-type transitions in alkali atoms trapped in a cold matrix of inert gases.

\bibitem{budker2014proposal}
Dmitry Budker, Peter~W Graham, Micah Ledbetter, Surjeet Rajendran, and Alexander~O Sushkov.
\newblock Proposal for a cosmic axion spin precession experiment (casper).
\newblock {\em Phys. Rev. X}, 4(2):021030, 2014.

\bibitem{kuchler2014novel}
Florian Kuchler, P~Fierlinger, and D~Wurm.
\newblock A novel approach to measure the electric dipole moment of the isotope 129-xe.
\newblock In {\em EPJ Web of Conferences}, volume~66, page 05011. EDP Sciences, 2014.

\bibitem{wu2019search}
Teng Wu, John~W Blanchard, Gary~P Centers, Nataniel~L Figueroa, Antoine Garcon, Peter~W Graham, Derek F~Jackson Kimball, Surjeet Rajendran, Yevgeny~V Stadnik, Alexander~O Sushkov, et~al.
\newblock Search for axionlike dark matter with a liquid-state nuclear spin comagnetometer.
\newblock {\em Physical review letters}, 122(19):191302, 2019.

\bibitem{leggett1977macroscopic}
AJ~Leggett.
\newblock Macroscopic parity nonconservation due to neutral currents?
\newblock {\em Physical Review Letters}, 39(10):587, 1977.

\bibitem{leggett1978macroscopic}
AJ~Leggett.
\newblock Macroscopic effect of p-and t-nonconserving interactions in ferroelectrics: A possible experiment?
\newblock {\em Physical Review Letters}, 41(8):586, 1978.

\bibitem{flambaum1992long}
VV~Flambaum.
\newblock Long-range parity-nonconserving interaction.
\newblock {\em Physical Review A}, 45(9):6174, 1992.

\bibitem{bouchiat2001atomic}
Marie-Anne Bouchiat and Claude Bouchiat.
\newblock An atomic linear stark shift violating p but not t arising from the electroweak nuclear anapole moment.
\newblock {\em The European Physical Journal D-Atomic, Molecular, Optical and Plasma Physics}, 15(1):5--18, 2001.

\bibitem{mukhamedjanov2005manifestations}
TN~Mukhamedjanov, OP~Sushkov, and JM~Cadogan.
\newblock Manifestations of nuclear anapole moments in solid-state nmr.
\newblock {\em Phys. Rev. A}, 71(1):012107, 2005.

\bibitem{singh2019new}
Jaideep~Taggart Singh.
\newblock A new concept for searching for time-reversal symmetry violation using pa-229 ions trapped in optical crystals.
\newblock {\em Hyperfine Interactions}, 240(1):29, 2019.

\bibitem{ramachandran2023nuclear}
Harish~D Ramachandran and Amar~C Vutha.
\newblock Nuclear t-violation search using octupole-deformed nuclei in a crystal.
\newblock {\em arXiv preprint arXiv:2304.10331}, 2023.

\bibitem{morris2023rare}
Ian~M Morris, Kai Klink, Jaideep~T Singh, Jose~L Mendoza-Cortes, Shannon~S Nicley, and Jonas~N Becker.
\newblock Rare isotope-containing diamond color centers for fundamental symmetry tests.
\newblock {\em arXiv preprint arXiv:2305.05781}, 2023.

\bibitem{sushkov2023effective}
AO~Sushkov, OP~Sushkov, and A~Yaresko.
\newblock Effective electric field: Quantifying the sensitivity of searches for new p, t-odd physics with eucl 3{\textperiodcentered} 6 h 2 o.
\newblock {\em Physical Review A}, 107(6):062823, 2023.

\bibitem{Andresen:2010jba}
G~B Andresen, M~D Ashkezari, M~Baquero-Ruiz, W~Bertsche, P~D Bowe, E~Butler, C~L Cesar, S~Chapman, M~Charlton, A~Deller, S~Eriksson, J~Fajans, T~Friesen, M~C Fujiwara, D~R Gill, A~Gutierrez, J~S Hangst, W~N Hardy, M~E Hayden, A~J Humphries, R~Hydomako, M~J Jenkins, S~Jonsell, L~V J{\o}rgensen, L~Kurchaninov, N~Madsen, S~Menary, P~Nolan, K~Olchanski, A~Olin, A~Povilus, P~Pusa, F~Robicheaux, E~Sarid, S~Seif~El Nasr, D~M Silveira, C~So, J~W Storey, R~I Thompson, D~P van~der Werf, J~S Wurtele, and Y~Yamazaki.
\newblock {Trapped antihydrogen}.
\newblock {\em Nature}, 468(7324):673--676, February 2010.

\bibitem{Ahmadi:2017a}
M.~Ahmadi, B.~X.~R. Alves, C.~J. Baker, W.~Bertsche, E.~Butler, A.~Capra, C.~Carruth, C.~L. Cesar, M.~Charlton, S.~Cohen, R.~Collister, S.~Eriksson, A.~Evans, N.~Evetts, J.~Fajans, T.~Friesen, M.~C. Fujiwara, D.~R. Gill, A.~Gutierrez, J.~S. Hangst, W.~N. Hardy, M.~E. Hayden, C.~A. Isaac, A.~Ishida, M.~A. Johnson, S.~A. Jones, S.~Jonsell, L.~Kurchaninov, N.~Madsen, M.~Mathers, D.~Maxwell, J.~T.~K. McKenna, S.~Menary, J.~M. Michan, T.~Momose, J.~J. Munich, P.~Nolan, K.~Olchanski, A.~Olin, P.~Pusa, C.~{\O}. Rasmussen, F.~Robicheaux, R.~L. Sacramento, M.~Sameed, E.~Sarid, D.~M. Silveira, S.~Stracka, G.~Stutter, C.~So, T.~D. Tharp, J.~E. Thompson, R.~I. Thompson, D.~P. van~der Werf, and J.~S. Wurtele.
\newblock Observation of the 1s-2s transition in trapped antihydrogen.
\newblock {\em Nature}, 541(7638):506--510, January 2017.

\bibitem{Amole:2012bh}
C.~Amole, M.~D. Ashkezari, M.~Baquero-Ruiz, W.~Bertsche, P.~D. Bowe, E.~Butler, A.~Capra, C.~L. Cesar, M.~Charlton, A.~Deller, P.~H. Donnan, S.~Eriksson, J.~Fajans, T.~Friesen, M.~C. Fujiwara, D.~R. Gill, A.~Gutierrez, J.~S. Hangst, W.~N. Hardy, M.~E. Hayden, A.~J. Humphries, C.~A. Isaac, S.~Jonsell, L.~Kurchaninov, A.~Little, N.~Madsen, J.~T.~K. McKenna, S.~Menary, S.~C. Napoli, P.~Nolan, K.~Olchanski, A.~Olin, P.~Pusa, C.~O. Rasmussen, F.~Robicheaux, E.~Sarid, C.~R. Shields, D.~M. Silveira, S.~Stracka, C.~So, R.~I. Thompson, D.~P. van~der Werf, and J.~S. Wurtele.
\newblock Resonant quantum transitions in trapped antihydrogen atoms.
\newblock {\em Nature}, 483(7390):439--443, 03 2012.

\bibitem{Ahmadi:2017}
M.~Ahmadi, B.~X.R.~R. Alves, C.~J. Baker, W.~Bertsche, E.~Butler, A.~Capra, C.~Carruth, C.~L. Cesar, M.~Charlton, S.~Cohen, R.~Collister, S.~Eriksson, A.~Evans, N.~Evetts, J.~Fajans, T.~Friesen, M.~C. Fujiwara, D.~R. Gill, A.~Gutierrez, J.~S. Hangst, W.~N. Hardy, M.~E. Hayden, C.~A. Isaac, A.~Ishida, M.~A. Johnson, S.~A. Jones, S.~Jonsell, L.~Kurchaninov, N.~Madsen, M.~Mathers, D.~Maxwell, J.~T.K.~K. McKenna, S.~Menary, J.~M. Michan, T.~Momose, J.~J. Munich, P.~Nolan, K.~Olchanski, A.~Olin, P.~Pusa, C.~{\O}. Rasmussen, F.~Robicheaux, R.~L. Sacramento, M.~Sameed, E.~Sarid, D.~M. Silveira, S.~Stracka, G.~Stutter, C.~So, T.~D. Tharp, J.~E. Thompson, R.~I. Thompson, D.~P. {Van Der Werf}, and J.~S. Wurtele.
\newblock {Observation of the hyperfine spectrum of antihydrogen}.
\newblock {\em Nature}, 548(7665):66--69, 2017.

\bibitem{Ahmadi:2020}
M.~Ahmadi, B.~X.~R. Alves, C.~J. Baker, W.~Bertsche, A.~Capra, C.~Carruth, C.~L. Cesar, M.~Charlton, S.~Cohen, R.~Collister, S.~Eriksson, A.~Evans, N.~Evetts, J.~Fajans, T.~Friesen, M.~C. Fujiwara, D.~R. Gill, P.~Granum, J.~S. Hangst, W.~N. Hardy, M.~E. Hayden, E.~D. Hunter, C.~A. Isaac, M.~A. Johnson, J.~M. Jones, S.~A. Jones, S.~Jonsell, A.~Khramov, P.~Knapp, L.~Kurchaninov, N.~Madsen, D.~Maxwell, J.~T.~K. McKenna, S.~Menary, J.~M. Michan, T.~Momose, J.~J. Munich, K.~Olchanski, A.~Olin, P.~Pusa, C.~{\O}. Rasmussen, F.~Robicheaux, R.~L. Sacramento, M.~Sameed, E.~Sarid, D.~M. Silveira, C.~So, D.~M. Starko, G.~Stutter, T.~D. Tharp, R.~I. Thompson, D.~P. van~der Werf, and J.~S. Wurtele.
\newblock Investigation of the fine structure of antihydrogen.
\newblock {\em Nature}, 578(7795):375--380, February 2020.

\bibitem{Ahmadi:2018b}
M.~Ahmadi, B.~X.~R. Alves, C.~J. Baker, W.~Bertsche, A.~Capra, C.~Carruth, C.~L. Cesar, M.~Charlton, S.~Cohen, R.~Collister, S.~Eriksson, A.~Evans, N.~Evetts, J.~Fajans, T.~Friesen, M.~C. Fujiwara, D.~R. Gill, J.~S. Hangst, W.~N. Hardy, M.~E. Hayden, E.~D. Hunter, C.~A. Isaac, M.~A. Johnson, J.~M. Jones, S.~A. Jones, S.~Jonsell, A.~Khramov, P.~Knapp, L.~Kurchaninov, N.~Madsen, D.~Maxwell, J.~T.~K. McKenna, S.~Menary, J.~M. Michan, T.~Momose, J.~J. Munich, K.~Olchanski, A.~Olin, P.~Pusa, C.~{\O}. Rasmussen, F.~Robicheaux, R.~L. Sacramento, M.~Sameed, E.~Sarid, D.~M. Silveira, D.~M. Starko, G.~Stutter, C.~So, T.~D. Tharp, R.~I. Thompson, D.~P. van~der Werf, and J.~S. Wurtele.
\newblock Observation of the {1S}–{2P} {Lyman}-$\alpha$ transition in antihydrogen.
\newblock {\em Nature}, 561(7722):211--215, September 2018.

\bibitem{Parthey:2011ys}
Christian~G. Parthey, Arthur Matveev, Janis Alnis, Birgitta Bernhardt, Axel Beyer, Ronald Holzwarth, Aliaksei Maistrou, Randolf Pohl, Katharina Predehl, Thomas Udem, Tobias Wilken, Nikolai Kolachevsky, Michel Abgrall, Daniele Rovera, Christophe Salomon, Philippe Laurent, and Theodor~W. H\"ansch.
\newblock Improved measurement of the hydrogen $1s\char21{}2s$ transition frequency.
\newblock {\em Phys. Rev. Lett.}, 107:203001, Nov 2011.

\bibitem{Hayden:22}
M. Hayden, priv. comm. (2022).

\bibitem{Bluhm:1999vq}
R~Bluhm, VA~Kosteleck{\'y}, and N~Russell.
\newblock {CPT and Lorentz tests in hydrogen and antihydrogen}.
\newblock {\em Physical Review Letters}, 82(11):2254--2257, 1999.

\bibitem{Widmann:04}
E.~Widmann, R.S. Hayano, M.~Hori, and T.~Yamazaki.
\newblock Measurement of the hyperfine structure of antihydrogen.
\newblock {\em Nuclear Instruments and Methods in Physics Research Section B: Beam Interactions with Materials and Atoms}, 214(0):31 -- 34, 2004.
\newblock Low Energy Antiproton Physics (LEAP'03).

\bibitem{Malbrunot:2017}
C.~Malbrunot, C.~Amsler, S.~Arguedas Cuendi, H.~Breuker, P.~Dupre, M.~Fleck, H.~Higaki, Y.~Kanai, T.~Kobayashi, B.~Kolbinger, N.~Kuroda, M.~Leali, V.~Maeckel, V.~Mascagna, O.~Massiczek, Y.~Matsuda, Y.~Nagata, M.C.~C. Simon, H.~Spitzer, M.~Tajima, S.~Ulmer, L.~Venturelli, E.~Widmann, M.~Wiesinger, Y.~Yamazaki, J.~Zmeskal, S.~Arguedas~Cuendi, H.~Breuker, P.~Dupre, M.~Fleck, H.~Higaki, Y.~Kanai, T.~Kobayashi, B.~Kolbinger, N.~Kuroda, M.~Leali, V.~Mäckel, V.~Mascagna, O.~Massiczek, Y.~Matsuda, Y.~Nagata, M.C.~C. Simon, H.~Spitzer, M.~Tajima, S.~Ulmer, L.~Venturelli, E.~Widmann, M.~Wiesinger, Y.~Yamazaki, and J.~Zmeskal.
\newblock The {ASACUSA} antihydrogen and hydrogen program : results and prospects.
\newblock {\em Phil. Trans. R. Soc.. A}, 376:20170273--20170273, 2017.

\bibitem{Widmann:2019}
E.~Widmann, C.~Amsler, S.~Arguedas~Cuendis, H.~Breuker, M.~Diermaier, P.~Dupré, C.~Evans, M.~Fleck, A.~Gligorova, H.~Higaki, Y.~Kanai, B.~Kolbinger, N.~Kuroda, M.~Leali, A.M.M. Leite, V.~Mäckel, C.~Malbrunot, V.~Mascagna, O.~Massiczek, Y.~Matsuda, D.J. Murtagh, Y.~Nagata, A.~Nanda, D.~Phan, C.~Sauerzopf, M.C. Simon, M.~Tajima, H.~Spitzer, M.~Strube, S.~Ulmer, L.~Venturelli, M.~Wiesinger, Y.~Yamazaki, and J.~Zmeskal.
\newblock Hyperfine spectroscopy of hydrogen and antihydrogen in {ASACUSA}.
\newblock {\em Hyperfine Interact.}, 240(1):5, 2019.

\bibitem{Diermaier:2017}
M.~Diermaier, C.~B. Jepsen, B.~Kolbinger, C.~Malbrunot, O.~Massiczek, C.~Sauerzopf, M.~C. Simon, J.~Zmeskal, and E.~Widmann.
\newblock In-beam measurement of the hydrogen hyperfine splitting and prospects for antihydrogen spectroscopy.
\newblock {\em Nature Communications}, 8(1):15749, 2017.

\bibitem{Ramsey:1949fk}
Norman~F. Ramsey.
\newblock A new molecular beam resonance method.
\newblock {\em Phys. Rev.}, 76:996--996, Oct 1949.

\bibitem{Ramsey:1950ud}
NF~Ramsey.
\newblock {A molecular beam resonance method with separated oscillating fields}.
\newblock {\em Physical Review}, 78:695--699, 1950.

\bibitem{BGross_1998}
B.~Gross, A.~Huber, M.~Niering, M.~Weitz, and T.~W. Hänsch.
\newblock Optical ramsey spectroscopy of atomic hydrogen.
\newblock {\em Europhysics Letters}, 44(2):186, oct 1998.

\bibitem{10.1063/5.0070037}
J~Ahokas, A~Semakin, J~J{\"a}rvinen, O~Hanski, A~Laptiyenko, V~Dvornichenko, K~Salonen, Z~Burkley, P~Crivelli, A~Golovizin, V~Nesvizhevsky, F~Nez, P~Yzombard, E~Widmann, and S~Vasiliev.
\newblock {A large octupole magnetic trap for research with atomic hydrogen}.
\newblock {\em Review of Scientific Instruments}, 93(2):023201, February 2022.

\bibitem{Kasevich:89}
Mark~A. Kasevich, Erling Riis, Steven Chu, and Ralph~G. DeVoe.
\newblock rf spectroscopy in an atomic fountain.
\newblock {\em Phys. Rev. Lett.}, 63:612--615, Aug 1989.

\bibitem{Wynands:2005}
R~Wynands and S~Weyers.
\newblock Atomic fountain clocks.
\newblock {\em Metrologia}, 42(3):S64, jun 2005.

\bibitem{Beausoleil86}
R.~G. Beausoleil and T.~W. H\"ansch.
\newblock Ultrahigh-resolution two-photon optical ramsey spectroscopy of an atomic fountain.
\newblock {\em Phys. Rev. A}, 33:1661--1670, Mar 1986.

\bibitem{Fried:98}
Dale~G. Fried, Thomas~C. Killian, Lorenz Willmann, David Landhuis, Stephen~C. Moss, Daniel Kleppner, and Thomas~J. Greytak.
\newblock Bose-einstein condensation of atomic hydrogen.
\newblock {\em Phys. Rev. Lett.}, 81:3811--3814, Nov 1998.

\bibitem{GRASIAN}
The grasian collaboration.
\newblock website: \url{https://grasian.eu}.

\bibitem{HAICU}
Makoto Fujiwara and Takamasa Momose.
\newblock private communications and talk at tipp2021, 2023.
\newblock Available online at: \url{https://indico.cern.ch/event/981823/contributions/4341752/attachments/2249762/3819375/TIPP2021_Sub3.pdf}, last accessed on 15.08.2023.

\bibitem{Breit28}
G.~Breit.
\newblock The propagation of schroedinger waves in a uniform field of force.
\newblock {\em Phys. Rev.}, 32:273--276, Aug 1928.

\bibitem{Goldman60}
V.~I.~Kogan I.~I.~Goldman, V. D.~Krivchenkov and V.~M. Galitski.
\newblock {\em Problems in Quantum Mechanics}.
\newblock Academic Press, New York,NY, USA, 1960.

\bibitem{Landau77}
E.M. Landau, L.D.;~Lifshitz.
\newblock {\em Quantum Mechanics}.
\newblock Pergamon Press: Oxford, 1977.

\bibitem{Nesvizhevsky:2002vv}
Valery~V Nesvizhevsky, Hans~G B{\"o}rner, Alexander~K Petukhov, Hartmut Abele, Stefan Bae{\ss}ler, Frank~J Rue{\ss}, Thilo St{\"o}ferle, Alexander Westphal, Alexei~M Gagarski, Guennady~A Petrov, and Alexander~V Strelkov.
\newblock {Quantum states of neutrons in the Earth's gravitational field}.
\newblock {\em Nature}, 415(6869):297--299, 2002.

\bibitem{Westphal_2007}
A.~Westphal, H.~Abele, S.~Bae{\ss}ler, V.V. Nesvizhevsky, K.V. Protasov, and A.Y. Voronin.
\newblock A quantum mechanical description of the experiment on the observation of gravitationally bound states.
\newblock {\em The European Physical Journal C}, 51(2):367--375, jun 2007.

\bibitem{ANTONIADIS2011}
I.~Antoniadis, S.~Baessler, M.~Büchner, V.V. Fedorov, S.~Hoedl, A.~Lambrecht, V.V. Nesvizhevsky, G.~Pignol, K.V. Protasov, S.~Reynaud, and Yu. Sobolev.
\newblock Short-range fundamental forces.
\newblock {\em Comptes Rendus Physique}, 12(8):755--778, 2011.
\newblock Ultra cold neutron quantum states.

\bibitem{OBertolami_2003}
O~Bertolami and F~M Nunes.
\newblock Ultracold neutrons, quantum effects of gravity and the weak equivalence principle.
\newblock {\em Classical and Quantum Gravity}, 20(5):L61, feb 2003.

\bibitem{Saha06}
A.~Saha.
\newblock Time-space non-commutativity in gravitational quantum well scenario.
\newblock {\em Eur. Phys. J. C}, 51:199--205, 2007.

\bibitem{Escobar2019b}
C.~A. Escobar and A.~Mart\'{\i}n-Ruiz.
\newblock Gravitational searches for lorentz violation with ultracold neutrons.
\newblock {\em Phys. Rev. D}, 99:075032, Apr 2019.

\bibitem{IVANOV2019134819}
A.N. Ivanov, M.~Wellenzohn, and H.~Abele.
\newblock Probing of violation of lorentz invariance by ultracold neutrons in the standard model extension.
\newblock {\em Physics Letters B}, 797:134819, 2019.

\bibitem{Crepin19_WEP}
P.-P. Cr\'epin, C.~Christen, R.~Gu\'erout, V.~V. Nesvizhevsky, A.Yu. Voronin, and S.~Reynaud.
\newblock Quantum interference test of the equivalence principle on antihydrogen.
\newblock {\em Phys. Rev. A}, 99:042119, Apr 2019.

\bibitem{Dufour13}
G.~Dufour, A.~G\'erardin, R.~Gu\'erout, A.~Lambrecht, V.~V. Nesvizhevsky, S.~Reynaud, and A.~Yu. Voronin.
\newblock Quantum reflection of antihydrogen from the casimir potential above matter slabs.
\newblock {\em Phys. Rev. A}, 87:012901, Jan 2013.

\bibitem{Kar05}
Savely~G. Karshenboim.
\newblock Precision physics of simple atoms: Qed tests, nuclear structure and fundamental constants.
\newblock {\em Physics Reports}, 422(1):1--63, 2005.

\bibitem{Buisseret_2007}
Fabien Buisseret, Bernard Silvestre-Brac, and Vincent Mathieu.
\newblock Modified newton's law, braneworlds, and the gravitational quantum well.
\newblock {\em Classical and Quantum Gravity}, 24(4):855, jan 2007.

\bibitem{Kosteleck15}
V.~Alan Kosteleck\'y and Arnaldo~J. Vargas.
\newblock Lorentz and $cpt$ tests with hydrogen, antihydrogen, and related systems.
\newblock {\em Phys. Rev. D}, 92:056002, Sep 2015.

\bibitem{GQSPs2015}
P.~Crivelli, , V.~V. Nesvizhevsky, and A.~Yu. Voronin.
\newblock Can we observe the gravitational quantum states of positronium?
\newblock {\em Advances in High Energy Physics}, 2015:173572, 2015.

\bibitem{Cooper2016}
B.~S. Cooper, A.~M. Alonso, A.~Deller, L.~Liszkay, and D.~B. Cassidy.
\newblock Positronium production in cryogenic environments.
\newblock {\em Phys. Rev. B}, 93:125305, Mar 2016.

\bibitem{Killian23}
C.~Killian, Z.~Burkley, P.~Blumer, et~al.
\newblock {T}owards the first demonstration of gravitational quantum states of atoms with a cryogenic hydrogen beam.
\newblock {\em Eur. Phys. J. D}, 77:50, 2023.

\bibitem{BERTOLAMI2006111}
O.~Bertolami and J.G. Rosa.
\newblock Quantum and classical divide: the gravitational case.
\newblock {\em Physics Letters B}, 633(1):111--115, 2006.

\bibitem{Voronin11}
A.~Yu. Voronin, P.~Froelich, and V.~V. Nesvizhevsky.
\newblock Gravitational quantum states of antihydrogen.
\newblock {\em Phys. Rev. A}, 83:032903, Mar 2011.

\bibitem{Berkhout89}
J.~J. Berkhout, O.~J. Luiten, I.~D. Setija, T.~W. Hijmans, T.~Mizusaki, and J.~T.~M. Walraven.
\newblock Quantum reflection: Focusing of hydrogen atoms with a concave mirror.
\newblock {\em Phys. Rev. Lett.}, 63:1689--1692, Oct 1989.

\bibitem{Doyle91}
John~M. Doyle, Jon~C. Sandberg, Ite~A. Yu, Claudio~L. Cesar, Daniel Kleppner, and Thomas~J. Greytak.
\newblock Hydrogen in the submillikelvin regime: Sticking probability on superfluid $^{4}\mathrm{He}$.
\newblock {\em Phys. Rev. Lett.}, 67:603--606, Jul 1991.

\bibitem{Berkhout93}
J.~J. Berkhout and J.~T.~M. Walraven.
\newblock Scattering of hydrogen atoms from liquid-helium surfaces.
\newblock {\em Phys. Rev. B}, 47:8886--8904, Apr 1993.

\bibitem{Yu93}
Ite~A. Yu, John~M. Doyle, Jon~C. Sandberg, Claudio~L. Cesar, Daniel Kleppner, and Thomas~J. Greytak.
\newblock Evidence for universal quantum reflection of hydrogen from liquid $^{4}\mathrm{He}$.
\newblock {\em Phys. Rev. Lett.}, 71:1589--1592, Sep 1993.

\bibitem{Nayak83}
V.~U. Nayak, D.~O. Edwards, and N.~Masuhara.
\newblock Scattering of $^{4}\mathrm{He}$ atoms grazing the liquid-$^{4}\mathrm{He}$ surface.
\newblock {\em Phys. Rev. Lett.}, 50:990--992, Mar 1983.

\bibitem{Shimizu01}
Fujio Shimizu.
\newblock Specular reflection of very slow metastable neon atoms from a solid surface.
\newblock {\em Phys. Rev. Lett.}, 86:987--990, Feb 2001.

\bibitem{Pasquini04}
T.~A. Pasquini, Y.~Shin, C.~Sanner, M.~Saba, A.~Schirotzek, D.~E. Pritchard, and W.~Ketterle.
\newblock Quantum reflection from a solid surface at normal incidence.
\newblock {\em Phys. Rev. Lett.}, 93:223201, Nov 2004.

\bibitem{Crepin17}
P.-P. Crépin, E.~A. Kupriyanova, R.~Guerout, A.~Lambrecht, V.~V. Nesvizhevsky, S.~Reynaud, S.~Vasiliev, and A.~Yu. Voronin.
\newblock Quantum reflection of antihydrogen from a liquid helium film.
\newblock {\em Europhysics Letters}, 119:3, 2017.

\bibitem{Crepin19_QR}
PP. Crépin, Kupriyanova, E.A., R.~Guérout, and et~al.
\newblock Quantum reflection of antihydrogen from a liquid helium bulk.
\newblock {\em Hyperfine Interact}, 240:58, 2019.

\bibitem{Voronin16}
A~Yu Voronin, V~V Nesvizhevsky, G~Dufour, and S~Reynaud.
\newblock Quantum ballistic experiment on antihydrogen fall.
\newblock {\em Journal of Physics B: Atomic, Molecular and Optical Physics}, 49(5):054001, feb 2016.

\bibitem{Indelicato14}
P.~Indelicato, G.~Chardin, Grandemange P., and et~al.
\newblock The gbar project, or how does antimatter fall?. hyperfine interact.
\newblock {\em Hyperfine Interact}, 228:141–150, 2014.

\bibitem{PhysRevA.95.032501}
P.-P. Cr\'epin, G.~Dufour, R.~Gu\'erout, A.~Lambrecht, and S.~Reynaud.
\newblock Casimir-polder shifts on quantum levitation states.
\newblock {\em Phys. Rev. A}, 95:032501, Mar 2017.

\bibitem{Rouselle22}
O.~Rousselle, Cladé, P., S.~Guellati-Khélifa, and et~al.
\newblock Quantum interference measurement of the free fall of anti-hydrogen.
\newblock {\em Eur. Phys. J. D}, 76:209, 2022.

\bibitem{ALPHA_firstg}
C.~Amole, M.~D. Ashkezari, M.~Baquero-Ruiz, W.~Bertsche, E.~Butler, A.~Capra, C.~L. Cesar, M.~Charlton, S.~Eriksson, J.~Fajans, T.~Friesen, M.~C. Fujiwara, D.~R. Gill, A.~Gutierrez, J.~S. Hangst, W.~N. Hardy, M.~E. Hayden, C.~A. Isaac, S.~Jonsell, L.~Kurchaninov, A.~Little, N.~Madsen, J.~T.~K. McKenna, S.~Menary, S.~C. Napoli, P.~Nolan, A.~Olin, P.~Pusa, C.~{\O} Rasmussen, F.~Robicheaux, E.~Sarid, D.~M. Silveira, C.~So, R.~I. Thompson, D.~P. van~der Werf, J.~S. Wurtele, A.~I. Zhmoginov, A.~E. Charman, and The~ALPHA Collaboration.
\newblock Description and first application of a new technique to measure the gravitational mass of antihydrogen.
\newblock {\em Nature Communications}, 4(1):1785, 2013.

\bibitem{Rouselle22_fewN}
Olivier Rousselle, Pierre Clad\'e, Sa\"{\i}da Guellati-Kh\'elifa, Romain Gu\'erout, and Serge Reynaud.
\newblock Improving the statistical analysis of antihydrogen free fall by using near-edge events.
\newblock {\em Phys. Rev. A}, 105:022821, Feb 2022.

\bibitem{Safonov98_BEC_Turku}
A.~I. Safonov, S.~A. Vasilyev, I.~S. Yasnikov, I.~I. Lukashevich, and S.~Jaakkola.
\newblock Observation of quasicondensate in two-dimensional atomic hydrogen.
\newblock {\em Phys. Rev. Lett.}, 81:4545--4548, Nov 1998.

\bibitem{Vasiliev19}
S.~Vasiliev, Ahokas J., J.~Järvinen, and et~al.
\newblock Gravitational and matter-wave spectroscopy of atomic hydrogen at ultra-low energies.
\newblock {\em Hyperfine Interact}, 240:14, 2019.

\bibitem{MGT_20}
V.V. Nesvizhevsky, Nez, Vasiliev F., S.A., and et~al.
\newblock A magneto-gravitational trap for studies of gravitational quantum states.
\newblock {\em Eur. Phys. J. C}, 80:520, 2020.

\bibitem{Lambrecht11}
S.~Reynaud A.~Lambrecht.
\newblock Casimir and short-range gravity tests.
\newblock {\em 46th Rencontres de Moriond on Gravitational Waves and Experimental Gravity, Italy}, pages 199--206, 2011.

\end{thebibliography}

\end{document}